\newif\ifdraft \drafttrue
\newif\iffull  \fulltrue

\makeatletter \@input{tex.flags} \makeatother

\documentclass[conference]{IEEEtran}

\newtheorem{definition}{Definition}

\newtheorem{theorem}{Theorem}
\newtheorem{lemma}{Lemma}

\usepackage{amsopn}
\usepackage{xspace}
\usepackage{infer}                % Inference rules
\usepackage{stmaryrd}             % Double brackets
\usepackage[procnames]{listings}
\usepackage[dvipsnames]{xcolor}
\usepackage{amsfonts}
\usepackage{amsmath}
\usepackage{dsfont}               % Double stroke symbols (for
\usepackage{mathtools}
\usepackage{mathpartir}

\usepackage{graphicx} %subfigure
\usepackage{subcaption} %subfigure
\usepackage[bookmarks=false]{hyperref}
\definecolor{DarkGreen}{rgb}{0.1,0.5,0.1}
\definecolor{DarkRed}{rgb}{0.5,0.1,0.1}
\definecolor{DarkBlue}{rgb}{0.1,0.1,0.6}
\definecolor{DarkYellow}{rgb}{0.6,0.3,0.0}
\hypersetup{
    colorlinks=true,       % false: boxed links; true: colored links
    linkcolor=DarkRed,          % color of internal links
    citecolor=DarkGreen,        % color of links to bibliography
    urlcolor=DarkYellow,          % color of external links
}

\newcounter{todos}

\providecommand{\eqdef}{\raisebox{-.2ex}[.2ex]{$\stackrel{\textrm{\tiny \normalfont{def}}}{~=~}$}}

\providecommand{\implies}{\Rightarrow}
\newcommand{\pWHILE}{\textsf{p}\textsc{While}\xspace}

\newcommand{\true}{\mathsf{true}}

\newcommand{\dom}{\mathsf{dom}}
\newcommand{\Skip}{\mathsf{skip}}
\newcommand{\Seq}[2]{{#1};\,{#2}}
\newcommand{\Ass}[2]{#1 \leftarrow #2}
\newcommand{\Rand}[2]{#1 \stackrel{\raisebox{-.25ex}[.25ex]%
 {\tiny $\mathdollar$}}{\raisebox{-.2ex}[.2ex]{$\leftarrow$}} #2}
\newcommand{\Cond}[3]{\mathsf{if}\ #1\ \mathsf{then}\ #2\ \mathsf{else}\ #3}

\newcommand{\While}[2]{\mathsf{while}\ #1\ \mathsf{do}\ #2}

\newcommand{\Return}{\mathsf{return}}
\newcommand{\kwdassert}{\mathsf{assert}}

\newcommand{\Assert}[1]{\kwdassert\,({#1})}

\let\exp\relax
\newcommand{\exp}[1]{\mathsf{exp}{\left({#1}\right)}}

\renewcommand{\Pr}[2]{\mathrm{Pr}\left[#1 : #2\right]}

\newcommand{\charfun}{\mathds{1}}
\newcommand{\distr}{\mathcal{D}}
\newcommand{\unit}{\mathsf{unit}}

\newcommand{\glift}[5]{{#2}\,{#1}_{\!\langle#4,#5\rangle}\,{#3}}
\newcommand{\sem}[1]{\llbracket #1 \rrbracket}

\newcommand{\subst}[2]{\left\{#2/#1\right\}}
\newcommand{\evalexpr}[1]{\sem{#1}_{\Expr}}

\newcommand{\Var}{\mathcal{V}}
\newcommand{\Expr}{\mathcal{E}}

\newcommand{\Cmd}{\mathcal{C}}

\newcommand{\Mem}{\mathcal{M}}

\newcommand{\AEquiv}[6]{\vdash {#2} \sim_{\!\left\langle#5,#6\right\rangle} {#3} : {#1} \Longrightarrow {#4}}
\newcommand{\HL}[3]{\vdash {#2} : {#1} \Longrightarrow {#3}}

\newcommand{\Pre}{\Psi}
\newcommand{\Post}{\Phi}
\newcommand{\Inv}{\Theta}
\newcommand{\side}[1]{\langle #1 \rangle}
\newcommand{\sidel}{\side{1}}
\newcommand{\sider}{\side{2}}
\newcommand{\gdistsymb}[1]{\Delta_{#1}}
\newcommand{\gdist}[3]{\gdistsymb{#1} \!\left( #2,\linebreak[1] #3\right)}

\newcommand{\pinq}{\textsf{PINQ}\xspace}

\newcommand{\SProd}[1]{\lceil{#1}\rceil}

\newcommand{\alphav}{{\tt v}_\epsilon} 
\newcommand{\deltav}{{\tt v}_\delta} 
\newcommand{\outv}{{\tt out}} 
 
\newcommand{\nm}[1]{{\sf #1}\,}
\newcommand{\PLap}{\mathsf{Lap}^{\diamond}}
\newcommand{\PExp}{\mathsf{Exp}^{\diamond}}

\title{Proving differential privacy in Hoare logic \vspace{-1ex}}
\author{
  \IEEEauthorblockN{Gilles Barthe\IEEEauthorrefmark{1},
    Marco Gaboardi\IEEEauthorrefmark{3}, 
    Emilio Jes\'{u}s Gallego Arias\IEEEauthorrefmark{2},
    Justin Hsu\IEEEauthorrefmark{2},
    C\'{e}sar Kunz\IEEEauthorrefmark{1},
    and
    Pierre-Yves Strub\IEEEauthorrefmark{1}
  }\\
  \IEEEauthorblockA{
    \IEEEauthorrefmark{1}IMDEA Software Institute, Spain
    \qquad \qquad
    \IEEEauthorrefmark{3}University of Dundee, Scotland
    \qquad \qquad
    \IEEEauthorrefmark{2}University of Pennsylvania, USA
  }
}

\lstnewenvironment{easycrypt}[2][]%
  {\lstset{language=easycrypt,caption={#2},#1}}%
  {}

\lstdefinelanguage{easycrypt}{
  style=easycrypt-default,
  keywordsprefix={'},
  morekeywords=[1]{unit,bool,int,real,bitstring,array,list,matrix,word},
  morekeywords=[2]{type,op,axiom,lemma,module,pred,const,declare},
  morekeywords=[3]{var,fun},
  morekeywords=[4]{while,if},
  morekeywords=[5]{theory,end,clone,import,export,as,with,section},
  morekeywords=[6]{forall,exists,lambda},
  morekeywords=[7]{idtac,change,beta,iota,zeta,logic,delta,simplify,congr,generalize,
                   pose,split,left,right,case,intros,cut,elim,apply,rewrite,elimT,subst,
                   progress,trivial},
  morekeywords=[8]{by,assumption,smt,reflexivity},
  morekeywords=[9]{first,last,do,try},
  morecomment=[n][\itshape]{(*}{*)},
  morecomment=[n][\bfseries]{(**}{*)}
}

\lstdefinestyle{easycrypt-default}{
  columns=fullflexible,
  captionpos=b,
  frame=tb,
  xleftmargin=.1\textwidth,
  xrightmargin=.1\textwidth,
  rangebeginprefix={(**\ begin\ },
  rangeendprefix={(**\ end\ },
  rangesuffix={\ *)},
  includerangemarker=false,
  basicstyle=\small\sffamily,
  identifierstyle={},
  keywordstyle=[1]{\itshape\color{OliveGreen}},
  keywordstyle=[2]{\bfseries\color{Blue}},
  keywordstyle=[3]{\bfseries},
  keywordstyle=[4]{\bfseries},
  keywordstyle=[5]{\bfseries\color{OliveGreen}},
  keywordstyle=[6]{\itshape\color{Blue}},
  keywordstyle=[7]{\color{Blue}},
  keywordstyle=[8]{\color{Red}},
  keywordstyle=[9]{\color{OliveGreen}},
  literate={phi}{{$\!\phi\,$}}1
           {phi1}{{$\!\phi_1$}}1
           {phi2}{{$\!\phi_2$}}1
           {phi3}{{$\!\phi_3$}}1
           {phin}{{$\!\phi_n$}}1
}

\lstdefinestyle{easycrypt-pretty}{
    basicstyle=\small\sffamily,
    literate={:=}{{$\mathrel{\gets}$}}1
              {<=}{{$\mathrel{\leq}$}}1
              {>=}{{$\mathrel{\geq}$}}1
              {<>}{{$\mathrel{\neq}$}}1
              {=\$}{{$\stackrel{\$}{\gets}$}}1
              {forall}{{$\forall$}}1
              {exists}{{$\exists$}}1
              {->}{{$\rightarrow\;$}}1
              {<-}{{$\leftarrow\;$}}1
              {<->}{{$\leftrightarrow\;$}}1
              {<=>}{{$\Leftrightarrow\;$}}1
              {=>}{{$\Rightarrow\;$}}1
              {==>}{{$\Rrightarrow\;$}}1
              {\/\\}{{$\wedge$}}1
              {\\\/}{{$\vee$}}1
              {.\[}{{[}}1
              {''ora}{{$\mathrel{||}$}}1 %needed for correct display in index
              {'a}{{\color{OliveGreen}$\alpha\,$}}1
              {'b}{{\color{OliveGreen}$\beta\,$}}1
              {'c}{{\color{OliveGreen}$\gamma\,$}}1
              {'t}{{\color{OliveGreen}$\tau\,$}}1
              {'x}{{\color{OliveGreen}$\chi\,$}}1
              {lambda}{{$\lambda\,$}}1
}

\begin{document}
\maketitle
\begin{abstract}
  \emph{Differential privacy} is a rigorous, worst-case notion of
  privacy-preserving computation. Informally, a probabilistic program is
  differentially private if the participation of a single individual in the
  input database has a limited effect on the program's distribution on outputs.
  More technically, differential privacy is a quantitative $2$-safety property
  that bounds the distance between the output distributions of a probabilistic
  program on adjacent inputs. Like many $2$-safety properties, differential
  privacy lies outside the scope of traditional verification techniques.
  Existing approaches to enforce privacy are based on intricate,
  non-conventional type systems, or customized relational logics.  These
  approaches are difficult to implement and often cumbersome to use.

  We present an alternative approach that verifies differential privacy by
  standard, non-relational reasoning on non-probabilistic programs.  Our
  approach transforms a probabilistic program into a
  non-probabilistic program which simulates two executions of the original
  program. We prove that if the target program is correct with respect to a
  Hoare specification, then the original probabilistic program is differentially
  private. We provide a variety of examples from the differential privacy
  literature to demonstrate the utility of our approach.  Finally, we compare
  our approach with existing verification techniques for privacy.
\end{abstract}

\section{Introduction}
Program verification provides a rich array of techniques and tools for analyzing
program properties. However, they typically reason
about single program executions or trace properties. In contrast,
many security properties---such as non-interference in information flow
systems---require reasoning about {\em multiple} program executions. These {\em
  hyperproperties}~\cite{ClarksonS08} encompass many standard security analyses,
and lie outside the scope of standard verification tools---to date, there is no
generally applicable method or tool for verifying hyperproperties.  Instead,
ad hoc enforcement methods based on type systems, customized program logics, and
finite state automata analyses have been applied to specific hyperproperties.
While these approaches are effective, their design and implementation often
require significant effort.

A promising alternative is to reduce verification of a hyperproperty of a
program $c$ to verification of a standard property of a transformed program
$T(c)$. For instance, {\em self-composition}~\cite{BartheDR04,DarvasHS05} is a
general method for reducing {\em 2-safety} properties of a program $c$---which
reason about two runs of $c$---to safety properties of the sequential
composition $c;c'$, where $c'$ is a renaming of $c$.  Self-composition is
sound, complete, and applies to many programming languages and
verification settings. For instance, it has been used to verify information flow
properties using standard deductive methods like Hoare logic. 

A close relative of self-composition is the {\em synchronized product}
construction~\cite{ZaksP08}. This transformation also produces a program which
emulates two executions of the original program, but while self-composition performs
the executions in sequence, synchronized products perform the executions in {\em
  lockstep}, dramatically simplifying the verification task for certain
properties.  This transformation is an instance of the more general class of
{\em product transformations}, as studied by Zaks and Pnueli~\cite{ZaksP08}, and
recently by Barthe et al.~\cite{BartheCK11,BartheCK13}.

While there has been much research on combining product constructions
and deductive verification to reason about 2-safety for deterministic
programs, this approach remains largely unexplored for {\em
  probabilistic} programs.
% \gb{although there is a large body of literature of products of probabilistic
%   Labelled Transition Systems; not sure to say it here, or in the related work
%   section}
% \mg{I would leave as it is right now.}
This is not for lack of interesting use cases---many security notions of
probabilistic computation are naturally $2$-safety properties.

\paragraph*{Verifying differential privacy}
In this paper, we consider on one such property: {\em differential privacy}, which
provides strong guarantees for privacy-preserving
probabilistic computation. Formally, a probabilistic program $c$ is $(\epsilon,
  \delta)$-differentially private with respect to $\epsilon> 0$,
$\delta\geq 0$, and a relation\footnote{We are here taking a
  generalization of Differential Privacy with respect to an arbitrary relation
  $\Phi$. The usual definition is obtained by considering an \emph{adjacency}
relation between databases.}
% \mg{This can be removed if you don't
%   like. I think however it is better to state it as soon as we use the
% non-standard definition.}
$\Phi$ on the initial memories of $c$
if for every two initial memories $m_1$ and $m_2$ related by $\Phi$, and every
subset $A$ of output memories,
$$\Pr{c,m_1}{A} \leq \mathrm{exp}(\epsilon)\ \Pr{c,m_2}{A} + \delta .$$
Here $\Pr{c,m}{A}$ denotes the probability of the output memory landing in $A$
according to distribution $\sem{c}~m$, where $\sem{c}$ maps an initial
memory $m$ to a distribution $\sem{c}~m$ of output memories. Since this
definition concerns two runs of the same probabilistic program, differential
privacy is a probabilistic $2$-safety property.

Differentially private algorithms are typically built from two constructions:
{\em private mechanisms}, which add probabilistic noise to their input, and {\em
  composition}, which combines differentially private operations into a single
one. This compositional behavior makes differential privacy an attractive target
for program verification efforts.

Existing methods for proving differential privacy have been based on type
systems, automata analyses, and customized program logics. For
instance, {\sf Fuzz}~\cite{ReedP10}, {\sf DFuzz}~\cite{GaboardiHHNP13}
and related systems~\cite{EignerM13} enforce differential privacy
using linear type systems. This approach is expressive enough to type
many examples, but it is currently limited to {\em pure} differential
privacy (where $\delta=0$), and cannot handle more advanced examples.
Alternatively, Tschantz et al.~\cite{Tschantz201161} consider a
verification technique based on I/O automata; again, this approach is
limited to pure differential privacy. Finally, {\sf
  CertiPriv}~\cite{BartheKOZ13} and {\sf
  EasyCrypt}~\cite{BartheDGKZ13} use an approximate relational Hoare
logic for probabilistic programs to verify differential privacy.  This
approach is very expressive and can accommodate approximate
differential privacy (when $\delta\neq 0$), but relies on a custom
and complex logic. For instance, ad hoc rules for loops are required for many
advanced examples. Finally, a common weakness of all of the above approaches is
that their implementation is non-trivial.

\paragraph*{Self-products for differential privacy}
To avoid these drawbacks, we investigate a new approach
where proving $(\epsilon,\delta)$-differential privacy of a program
$c$ is reduced to proving a safety property of a transformed program
$T(c)$. In view of previous work verifying $2$-safety properties, a
natural choice for $T$ is some notion of product program. However, the transformed
programs would then be probabilistic, and there are few tools for deductive
verification of probabilistic programs.  Targeting a non-probabilistic
language is more appealing in this regard, as there are many established
tools for deductive verification of non-probabilistic programs. Since the
original program is a probabilistic program, a key part of our approach
is to remove the probabilistic behavior from the target.

To define the transformation, we proceed in two steps. Starting from a
probabilistic program $c$, we first construct the synchronized product of $c$
with itself. Using the synchronized product instead of self-composition is
essential for our second step, in which the probabilistic product program is
transformed into a non-probabilistic program.

For this step, we rely on specific features of the differential
privacy property. First, we observe that differential privacy bounds the
ratio---hereafter called the {\em privacy cost}---between the
probabilities of producing the same output on two executions on nearby
databases. Second, we recall that there are two main tools for building
differentially private computations: private mechanisms, and
composition. Private mechanisms and composition interact with the
privacy cost in different ways; we consider each in turn.

A private mechanism run over two different inputs returns two closely
related distributions, at the cost of consuming some privacy budget.
The privacy cost depends on the distance between the inputs: as the
two inputs become farther apart, the privacy cost also grows. One
fundamental insight (due to Barthe et al.~\cite{BartheKOZ12}) we use is
that the property of being \lq\lq closely related\rq\rq\ can be
understood as being at distance 0 for a suitable notion of distance on
distributions.

Composition takes a set of differentially private operations and returns the
sequential composition of the operations, which is also differentially private.
By a property of differential privacy, the privacy cost of the
composition is upper bounded by the sum of the privacy costs of the individual
operations.

We build this reasoning directly into our verification system. First,
we apply the synchronized product construction. Then, we
replace two corresponding calls to a mechanism with a call to an
abstract procedure that returns equal outputs, at the cost of
consuming some privacy budget---roughly, being at distance 0 in
the probabilistic setting is equivalent to being equal in the
non-probabilistic setting.  To keep track of the privacy cost, we use ghost
variables $\alphav$ and $\deltav$ which are incremented after each mechanism is
executed, in terms of the distance between their two inputs. 

Note that the second step leverages the synchronized product construction: since
the two executions are simulated in lockstep, corresponding calls to a mechanism
are next to each other in the product program.  Since mechanisms are the only
probabilistic parts of our source program, our output program is now
non-probabilistic.

To illustrate our approach, consider the following simple program $c$:
\begin{displaymath}
s;\ \Ass{x}{\mathsf{Lap}_\epsilon(e)};\ \mathsf{return}~x
\end{displaymath}
where $s$ is a deterministic computation and $\mathsf{Lap}$ is the
Laplace mechanism---a probabilistic operator that achieves
differential privacy by adding noise to its input. The synchronized product
$T(c)$ of the program $c$ is
\begin{displaymath}
T(s);\ 
\Ass{x_1}{\mathsf{Lap}_\epsilon(e_1)};\ 
\Ass{x_2}{\mathsf{Lap}_\epsilon(e_2)};\ 
\mathsf{return}~(x_1,x_2)
\end{displaymath}
where $T(s)$ is the synchronized product of $s$. Then, we
make the program non-probabilistic by replacing the two calls to the
Laplace mechanism  with a call to an
abstract procedure $\PLap$, giving the following transformed program $T(c)$.
%\jh{Are we missing the contract here?}
%
\begin{displaymath}
    T(s);
    \Ass{(x_1,x_2)}{\PLap(e_1,e_2)};
    % \Havoc{x_1}; \Ass{x_2}{x_1};
    % \alphav= \alphav + \epsilon|e_1 - e_2|;
    \mathsf{return}~(x_1,x_2) 
\end{displaymath}
% \begin{displaymath}
%   p \eqdef 
%   \begin{array}[t]{l}
%     T(c); \\
%     \Havoc{x_1};\ \Ass{x_2}{x_1}; \\
%     \alphav= \alphav + \epsilon|e_1 - e_2|;\\ 
%     \mathsf{return}~(x_1,x_2) 
%   \end{array}
% \end{displaymath}
% 
Roughly, the specification of the procedure invocation $\PLap$ states
that the same value is assigned to $x_1$ and $x_2$. Also, as side
effect, the variable $\alphav$ is updated to increment the privacy
cost, which depends on the distance between the inputs ($e_1,e_2$) to
the Laplace mechanism.

Our main result (Theorem~\ref{thm:hldiffpriv} in \S\ref{sec:self}) states that
once we perform this transformation, we can use plain Hoare logic to complete
the verification. More concretely, for the example above, we represent the
relation on memories $\Phi$ as a predicate $\hat{\Phi}$ on pairs of memories,
and prove that $c$ is $(\epsilon,0)$-differentially private if the following
Hoare specification is valid.
$$\HL{\hat{\Phi} \wedge \alphav=0}{T(c)}{ x_1 = x_2 \wedge \alphav\leq
  \epsilon_0}$$ 
In the remainder of this article we use the same representation for a relation
and its representation as a predicate on memories.

%% In the example, the validity of the transformed program $T(c)$ with
%% respect to the Hoare specification
%% %
%% 
%% %
%% entails that the original program $c$ is
%% $(\epsilon_0,0)$-differentially private with respect to
%% $\Phi$.\footnote{
%%   In fact, we will show that the validity of the above specification follows
%%   from the following Hoare specification:
%%   %
%%   $$\{ \Phi \} ~T(s)~ \{|e_1 - e_2| \leq \frac{\epsilon_0}{\epsilon} \}$$
%%   %
%%   Readers familiar with differential privacy can think of this judgment as
%%   bounding the {\em sensitivity} of expression $e$ by $\epsilon_0/\epsilon$.

\paragraph*{Contributions}
The main contribution of the paper (\S\ref{sec:self}) is a program
transformation that operates on programs built from sequential,
non-probabilistic constructs and differentially private, probabilistic
primitives---such as the Laplace and Exponential mechanisms.  The
transformed program is non-probabilistic, and differential privacy
of the original program can be reduced to a safety property of the
transformed program. Then we show in \S\ref{sec:aprhl} that our
approach subsumes the core {\sf apRHL} logic of Barthe et
al.~\cite{BartheKOZ13}, in the sense that every algorithm provable
with core {\sf apRHL} is also provable with our approach.

We illustrate the expressiveness of our approach in
\S\ref{sec:examples} by verifying differential privacy of several
probabilistic algorithms, including a recent algorithm that produces
synthetic datasets using a combination of the multiplicative weights
update rule and the exponential mechanism~\cite{HardtR10,HardtLM12},
and the Propose-Test-Release (PTR)
framework~\cite{DworkL09,ThakurtaS13}, which achieves approximate
differential privacy without relying on output perturbation. Finally,
we discuss the example of vertex cover, which is provable {\sf apRHL},
but cannot be handled directly by our approach.
% Finally in ~\S\ref{sec:pinq}, we relate our
% approach to the runtime verification approach taken by {\sf
%   PINQ}~\cite{McSherryM10}, where a monitor dynamically tracks the
% privacy budget consumed by a program, and aborts when the budget is
% exhausted.

\section{A primer on differential privacy}
Let us begin by recalling the basic definitions of differential privacy.
\begin{definition}
Let $\epsilon,\delta \geq 0$, and let $\Phi \subseteq \mathcal{S}
\times \mathcal{S}$ be a relation on $\mathcal{S}$. A randomized algorithm $K$
taking inputs in $\mathcal{S}$ and returning outputs in $\mathcal{R}$ is
{\em $(\epsilon, \delta)$-differentially private with respect to  $\Phi$}
% \mg{is there any specific reason why we take $\epsilon$ strictly greater than
%   $0$?}
% \jh{No, the definition works fine for $0$ too (but it's kind of trivial; the
%   program will ignore all inputs).}
% \mg{These theorists, they really miss formalism ;-). I changed it because later
%   on we consider both $(0,\delta)$-DP and $(\epsilon,0)$-DP. Clearly a minor
%   point but I prefer to have a definition where the notions we consider fit. I
%   also removed $\delta<1$ because I think the definition is still correct even
%   if then practically it doesn't make sense. Anyway, it is really a minor
%   point.}
% \jh{Right, I'd forgotten about that. Close enough!}
if for every two inputs $s_1,s_2\in \mathcal{S}$ such that $s_1\ \Phi\
s_2$ and every subset of outputs $A\subseteq \mathcal{R}$,
$$\Pr{K(s_1)}{A} \leq \mathrm{exp}(\epsilon)\ \Pr{K(s_2)}{A} + \delta .$$
When $\delta = 0$, we will call this {\em $\epsilon$-differential privacy}.
\end{definition}

Our definition is a variant of the original definition of differential privacy,
\cite{DMNS06} where input memories are considered to be databases and
$\Phi$ relates databases that differ in a single individual's data;
let us briefly explain the intuition of differential privacy in this setting.
Recall that differential privacy aims to conceal the participation of
individuals in a study. To distinguish between the participation or
non-participation of an individual, we think of two databases $D$ and $D'$ are
\emph{adjacent} or \emph{neighboring} if they differ only in the presence or
absence of a single record; note that the adjacency relation is necessarily
symmetric.

Differential privacy then states that the two distributions output by $K$ on
a pair of adjacent databases are close. In the simple case where $\delta=0$, the
definition above requires that the probability of any output changes by at most
a $\exp{\epsilon}$ factor when moving from one input to an adjacent input.  When
$\delta>0$ these bounds are still valid except with probability $\delta$. In
other words, $\epsilon$ controls the strength of the privacy bound, and $\delta$
is the probability of failure in ensuring the privacy bound.

\paragraph*{Building private programs}
Let $F$ be a deterministic computation with inputs in $\mathcal{T}$
and outputs in $\mathcal{R}$. Suppose that we want to make the
computation of $F$ $(\epsilon,\delta)$-differentially private with
respect to some relation $\Phi$. A natural way to achieve this goal is
to add random noise to the evaluation of $F$ on an input. In
general, the noise that we need to add depends not only on the
$\epsilon$ and $\delta$ parameters (which control the strength of the
privacy guarantee), but also on the {\em sensitivity} of $F$,
a quantity that is closely related to Lipschitz continuity for functions.
\begin{definition}
Assume that $F$ is real-valued, i.e.\, $\mathcal{R}=\mathbb{R}$, and
let $k>0$. We say that $F$ is \emph{$k$-sensitive with respect to
$\Phi$} if $|F(t_1) - F(t_2)| \leq k$ for all $t_1,t_2\in \mathcal{T}$
such that $t_1\ \Phi\ t_2$.
\end{definition}

A typical mechanism for privately releasing a $k$-sensitive function is the
\emph{Laplace mechanism}.
\begin{theorem}[\cite{Dwork06}]
\label{thm:laplace}
Suppose $\epsilon> 0$. The \emph{Laplace mechanism} is defined by
$$ 
\mathrm{Lap}_{\epsilon}(t) = t + v,
$$
where $v$ is drawn from the Laplace distribution $\mathcal{L}(1/\epsilon)$, i.e.
with probability density function
\[
  P(v) = \exp{-\epsilon|v|}.
\]
If $F$ is $k$-sensitive with respect to $\Phi$, then the probabilistic function
that maps $t$ to $\mathrm{Lap}_{\epsilon}(F(t))$ is
$(k\epsilon,0)$-differentially private with respect to $\Phi$.
\end{theorem}
%\jh{Better in PTR, I think.}

Additionally, the Laplace mechanism satisfies a simple accuracy bound.
\begin{lemma}\label{lem:accuracy}
Let $\epsilon,\delta>0$ and let $T=\log(2/\delta)/(2\epsilon)$. Then for
every $x$, ${\sf Lap}_{\epsilon}( x)\in (x-T, x+T)$ with probability
at least $1-\delta$.
\end{lemma}

%% The Laplace mechanism cannot be directly used to obtain an approximate
%% differentially private query, i.e., an $(\epsilon,\delta)$-differentially
%% private query for $\delta>0$. A mechanism that can be used to achieve
%% approximate differential privacy is instead the \emph{Gaussian mechanism}:
%% \begin{theorem}[\cite{Dwork06b}]
%% \label{thm:gauss}
%%   Suppose $\epsilon,c > 0$. Given a $c$-sensitive query $q$, the \emph{Gaussian mechanism} is
%%   defined by
%% $$
%%   \mathrm{Gau}_{c, \epsilon,\delta}(D) = q(D) + v,
%% $$
%%   where $v$ is drawn from the Gaussian distribution $\mathcal{N}(0,
%%   \sigma^2)$ with deviation
%%   \[
%%     \sigma=\frac{c \sqrt{2\ln(2/\delta)} }{\epsilon}.
%%   \]
%%   This mechanism is $(\epsilon,\delta)$-differentially private.
%% \end{theorem}
Another mechanism that is fundamental for differential privacy is the
\emph{Exponential mechanism}~\cite{McSherryT07}. Let $\mathcal{T}$ be the set of
inputs, typically thought of as the private information.  Let $\mathcal{R}$ be
a set of outputs, and consider  a function $F : \mathcal{T} \times \mathcal{R}
\rightarrow \mathbb{R}$, typically called the {\em score function}. 
We first extend the definition of sensitivity to this function.
\begin{definition}
Assume $F : \mathcal{T} \times \mathcal{R}
\rightarrow \mathbb{R}$  and
let $c>0$. We say that $F$ is \emph{$k$-sensitive on $\mathcal{T}$ with respect to
$\Phi$} if $|F(t_1,r) - F(t_2,r)| \leq k$ for all $t_1,t_2\in \mathcal{T}$
such that $t_1\ \Phi\ t_2$ and $r\in\mathcal{R}$.
\end{definition}

Then, the Exponential mechanism can be used
to output an element of $\mathcal{R}$ that approximately maximizes the score
function, if the score function is $k$-sensitive.
\begin{theorem}[\cite{McSherryT07}] \label{thm:exponential}
  Let $\epsilon,c > 0$.  Suppose that $F$ is $k$-sensitive in $\mathcal{T}$ with
  respect to $\Phi$.  The \emph{Exponential mechanism}\footnote{The Exponential
    mechanism as first introduced by McSherry and Talwar~\cite{McSherryT07} is
    parameterized by a prior distribution $\mu$ on $\mathcal{R}$. We consider
    the special case where $\mu$ is uniform; this suffices for typical
    applications. }
  $\mathrm{Exp}_{\epsilon}(F,t)$ takes as input $t \in \mathcal{T}$, and
  returns $r\in \mathcal{R}$ with probability equal to
  \[
    \frac{\exp{\epsilon F(t,r)/2}}{\sum_{r'\in\mathcal{R}}\exp{\epsilon
        F(t,r')/2}}.
  \]
  This mechanism is $(k\epsilon,0)$-differentially private with respect to
  $\Phi$.
\end{theorem}

A powerful feature of differential privacy is that by composing differentially
private mechanisms, we can construct new mechanisms that satisfy differential
privacy. However, the privacy guarantee will degrade: more operations on a
database will lead to more privacy loss.  In light of this composition property,
we will often think of the privacy parameters $\epsilon$ and $\delta$ of a
program as \emph{privacy budgets} that are consumed by sub-operations.  This is
formalized by the following composition theorem.
\begin{theorem}[\cite{mcsherry.pinq09}]
  \label{thm:compose}
Let $q_1$ be a $(\epsilon_1,\delta_1)$-differentially private query and let
$q_2$ be a $(\epsilon_2,\delta_2)$-differentially private query.  Then, their
composition $q(t)=(q_1(t),q_2(t))$ is
$(\epsilon_1+\epsilon_2,\delta_1+\delta_2)$-differentially private.
\end{theorem}

Finally, differential privacy is closed under {\em post-processing}---an output
of a private algorithm can be arbitrarily transformed, so long as this
processing does not involve the private database.
\begin{theorem}
  \label{thm:postprocess}
  \newcommand{\cR}{R}
  \renewcommand{\circle}{\circ}
  Let $q$ be $(\epsilon, \delta)$-differentially private mapping
  databases to some output range $\cR$, and let $f : \cR \rightarrow \cR'$ be an
  arbitrary function. Then, the post-processing $f \circle q$ is also
  $(\epsilon,\delta)$-differentially private.
\end{theorem}
% We conclude this section by providing an alternative characterization
% of $(\epsilon,\delta)$-differential privacy based on the notion of
% $\epsilon$-distance. This notion is adapted from the asymmetric notion
% of distance used in Barthe et al.~\cite{BartheKOZ13}.
% \mg{I think it is ok to introduce the $\epsilon$-distance here but we are using
%   distribution notation that is introduced only later. Perhaps the best is to
%   postpone or to use a more general notation.}
% \jh{Decided to move to appendix.}
% \begin{definition}[$\epsilon$-distance]\label{def:alpha-dist}
%   The $\epsilon$-distance $\Delta_\epsilon$ is defined as
%   % 
%   $$ 
%   \gdist{\epsilon}{\mu_1}{\mu_2} \eqdef 
%   \max_{S \subseteq A}\,(\mu_1\,S - \exp{\epsilon}\,\mu_2\,S),
%   $$
%   where $\mu\,S \eqdef \sum_{a \in S} \mu\,a$. Note that we define max over an
%   empty set to be $0$, so $\Delta_\epsilon(\mu_1, \mu_2) \geq 0$.
% \end{definition}
% It follows from the definition of $\epsilon$-distance that a
% probabilistic program $c$ is $(\epsilon, \delta)$-differentially
% private with respect to $\epsilon> 0$, $\delta\geq 0$, and a relation
% $\Phi$ on the initial memories of $c$ if for every two memories $m_1$
% and $m_2$ related by $\Phi$, we have 
% $$\gdist{\epsilon}{{\sem{c}~m_1}}{{\sem{c}~m_2}}\leq \delta.$$

\section{Self-products} \label{sec:self}
In this section, we formalize the verification of differential-privacy
using traditional Hoare logic. We start with some preliminary
definitions and the \pWHILE programming language, which will serve as
our source language.  Then, given a \emph{probabilistic} \pWHILE program $c$, we
show how to build a
\emph{non-probabilistic} program $T(c)$ that simulates two
  executions of $c$ on different inputs and tracks
the privacy cost via two ghost variables $\alphav$ and $\deltav$. We
show that the verification of $T(c)$ with respect to a particular Hoare logic
specification ensures differential privacy of the original program $c$.

\subsection{Distributions}
We define the set $\distr(A)$ of {\em sub-distributions} over a set $A$ as
the set of functions $\mu:A\to [0,1]$ with discrete
$\mathsf{support}(\mu)=\{x \mid \mu\,x\neq 0 \}$, such that $\sum_{x\in
  A}\mu\,x\leq 1$; when equality holds, $\mu$ is a true {\em distribution}. (We
will often refer to sub-distributions as distributions when
there is no confusion.) Sub-distributions can be given the structure of a
complete partial order: for all $\mu_1,\mu_2\in\mathcal{D}(A)$,
\begin{displaymath}
\mu_1\sqsubseteq\mu_2 \eqdef \forall a\in A.~ \mu_1\,a\leq \mu_2\,a.
\end{displaymath}

Moreover, sub-distributions can be given the structure of a monad: for any
function $g:A\to\distr(B)$ and distribution $\mu:\distr(A)$, we define
$g^\star\,\mu:\distr(B)$ to be the sub-distribution
$$ 
 g^\star\,\mu\,(b) \eqdef \sum_{a\in A}(g\,a\,b)(\mu\,a),
$$
for every $b \in B$.  Given an element $a\in A$, let $\charfun_a$ be the
probability distribution that assigns all mass to the value $a$.

We will use a normalization construction $(\cdot)^\#$ that takes as input a
function $f:B \to \mathbb{R}^{\geq 0}$ over a discrete set $B$ and returns
$(f)^\# \in \distr(B)$ such that the probability mass of $f^\#$ at $b$ is given
by
$$
(f)^\#\, b  \eqdef \frac{f\, b}{\sum_{b'\in B} f\, b' }.
$$
Intuitively, sampling from the distribution $(f)^\#$ is equivalent to sampling
``with probability proportional to'' $f$.

% The proof of our main theorem relies on a {\em lifting} operator that
% turns a relation on memories into a relation on memory distributions. 
% Given a relation on memories $\Post$, and real values
% $\epsilon,\delta$ we define the lifted relation on memory
% distributions $\glift{\Post}{}{}{\epsilon}{\delta}$ as follows.
% \begin{definition}
%   For all memory distributions $\mu_1,\mu_2$,
%   $\glift{\Post}{\mu_1}{\mu_2}{\epsilon}{\delta}$ iff there exists $\mu$ such
%   that:
%   \begin{enumerate}
%   \item $\pi_i\,\mu\leq\mu_i$,
%   \item $\forall m, \mu\,m\neq0\Rightarrow \Post\,m$, and
%   \item $\Delta_\epsilon(\mu_i,\pi_i\,\mu)\leq\delta$,
%   \end{enumerate}
%   where 
%   \begin{itemize}
%     \item $(\pi_1\,\mu)\,m_1=\sum_{m_2\in\Mem}\mu\,(m_1,m_2)$, and 
%     \item $(\pi_2\,\mu)\,m_2=\sum_{m_1\in\Mem}\mu\,(m_1,m_2)$.
%   \end{itemize}
% \end{definition}
% Notice that $\epsilon$-distance between distributions is closely related to the
% lifting of the equality relation, i.e.,\,
% \begin{equation}
% \glift{=}{\mu_1}{\mu_2}{\epsilon}{\delta}\qquad  \text{iff}\qquad
% \gdist{\epsilon}{\mu_1}{\mu_2}\leq \delta.
% \end{equation}
% Note that the second equation is precisely the condition on output distributions
% needed for $(\epsilon, \delta)$-differential privacy.
% \mg{Is this the main observation, right? Maybe we can state it as a
%   remark that we can refer later when we introduce the
%   self-product. For the moment I just put an equation.}
% \jh{Will move to appendix.}

\subsection{\pWHILE Language}
\pWHILE programs will serve as our source language, and are defined by the
following grammar:
\begin{displaymath}
\begin{array}{r@{\ \ }l@{\quad}l}
\Cmd ::= & \Skip \\
     \mid& \Seq{\Cmd}{\Cmd}        & \mbox{sequencing}\\
     \mid& \Ass{\Var}{\Expr}        & \mbox{deterministic assignment}\\
%    \mid& \Rand{\Var}{\DExpr}      & \mbox{random assignment}\\
     \mid& \Rand{\Var}{\mathsf{Lap}_\epsilon(\Expr)}      &
     \mbox{Laplace assignment}\\
     % \mid& \Rand{\Var}{\mathsf{Gau}_{\epsilon,\delta}(\Expr)}      &
     % \mbox{Gauss assignment}\\
     \mid& \Rand{\Var}{\mathsf{Exp}_{\epsilon}(\Expr,\Expr)}      &
     \mbox{Exponential assignment}\\
     \mid& \Cond{\Expr}{\Cmd}{\Cmd} & \mbox{conditional}\\
     \mid& \While{\Expr}{\Cmd}      & \mbox{while loop}\\
%    \mid& \Call{\Var}{\Proc}{\Expr,\ldots,\Expr} & \mbox{procedure
%     call}\\[4pt]
      \mid& \Return~\Expr      & \mbox{return expression}\\
%MG: Removed the procedure call
\end{array}
\end{displaymath}
Here, $\Var$ is a set of \emph{variables} and $\Expr$ is a set of
\emph{expressions}. We consider expressions including simply typed lambda terms
and basic operations on booleans, lists and integers. (\pWHILE is equipped with
a standard type system; we omit the typing rules.)

The probabilistic assignments involving $\mathsf{Lap}_\epsilon(\Expr)$ and
$\mathsf{Exp}_{\epsilon}(\Expr,\Expr)$ internalize the (discrete version of the)
mechanisms of Theorem~\ref{thm:laplace} and Theorem~\ref{thm:exponential}
respectively.
%\jh{Would be smoother if we could name a system it is similar to. STLC?  PCF?
%  Something else?}%
% \mg{I don't think it is needed, we mention right above that we have
% the simply typed LC in expressions.}
Note that for examples based on the exponential
mechanism we allow function types for representing the score
functions; alternatively these score functions can be modeled as
finite maps if their domain is finite (as will be the case in our examples).

The semantics of a well-typed \pWHILE program is defined by its
(probabilistic) action on {\em memories}; we denote the set of
memories by $\mathcal{M}$. A program memory $m\in\mathcal{M}$ is a
partial assignment of values to variables. Formally, the semantics of
a return-free \pWHILE program $c$ is a function
$\sem{c}:\mathcal{M}\to \mathcal{D}(\mathcal{M})$ mapping a memory
$m\in\mathcal{M}$ to a distribution $\sem{c}\,m\in
\mathcal{D}(\mathcal{M})$, as defined in
Fig.~\ref{fig:semantics}.

For simplicity, we only consider programs of the form $c;\Return~{e}$ in the
rest of this paper.  Then, the semantics of a program $c;\Return~{e}$ is simply
defined as
\[
  \sem{\Seq{c}{\Return~{e}}}\; m  \eqdef \lambda u.\, \charfun_{\sem{e}
    u}^\star\,(\sem{c}\,m).
\]

\begin{figure*}
\begin{displaymath}
\begin{array}{l@{\;\,}c@{\;\,}l}
\\[-1.2ex]
\sem{\Skip}\; m         &=& \charfun_{m} \\[1ex]
\sem{\Seq{c_1}{c_2}}\; m  &=& \sem{c_2}^\star\,(\sem{c_1}\,m) \\[1ex]
\sem{\Ass{x}{e}}\; m    &=& \charfun_{m\subst{x}{\evalexpr{e}\,m}} \\[1ex]
%
% \sem{\Rand{x}{\mathsf{Lap}_\epsilon(e)}}\; m   &=& 
% \left (\lambda v.\ \charfun_{m\subst{x}{v}}\right )^\star \left (\lambda
% b. \frac{\exp{ - \frac{\epsilon|b-\sem{e}\, m|}{2} }}{\sum_{b'} \exp{
%     -\frac{\epsilon|b'-\sem{e}\, m|}{2} }}\right )
% \\[8pt]
\sem{\Rand{x}{\mathsf{Lap}_\epsilon(e)}}\; m   &=& 
\left (\lambda v.\ \charfun_{m\subst{x}{v}}\right )^\star \left (\lambda
r. \exp{ - \frac{\epsilon|r-\sem{e}\, m|}{2} }\right )^\#
\\[8pt]
%
% \sem{\Rand{x}{\mathsf{Gau}_{\epsilon,\delta}(e)}}\; m   &=& 
%  (\lambda v.\ \charfun_{m\subst{x}{v}})^\star (\lambda
%  b. \frac{\epsilon}{2 \sqrt{\ln(2/\delta)
%      \pi}}
% \mathsf{exp}
% ({-\epsilon^2 (b -\sem{e}m)^2}/{4\ln(2/\delta)}))
% \\[5pt]
%
\sem{\Rand{x}{\mathsf{Exp}_{\epsilon}(s,e)}}\; m   &=& 
\left (\lambda v.\ \charfun_{m\subst{x}{v}} \right )^\star \left (\lambda
 r. 
 \exp{\frac{\epsilon
       \sem{s}m(\sem{e}m,r)}{2}}
\right )^\#
\\[8pt]
% \sem{\Rand{x}{\mathsf{Exp}_{\epsilon}(s,e)}}\; m   &=& 
% \left (\lambda v.\ \charfun_{m\subst{x}{v}} \right )^\star \left (\lambda
%  r. 
%  \frac{\exp{\frac{\epsilon
%        \sem{s}m(\sem{e}m,r)}{2}}}{\sum_{r'\in\mathcal{R}}\exp{\frac{\epsilon \sem{s}m(\sem{e}m,r')}{2}}}
% \right )
% \\[8pt]
%
\sem{\Cond{e}{c_1}{c_2}}\ m &=& \mathbf{if}\;(\evalexpr{e}\,m=\true)
  ~\mathbf{then}~(\sem{c_1}\,m)
  ~\mathbf{else}~(\sem{c_2}\,m)\\[1ex]
\sem{\While{e}{c}}\ m &=&
\bigsqcup \: w_i \, m\\[-6pt]
\end{array}
\end{displaymath}
\begin{displaymath}
\qquad \mbox{where }  \; 
\begin{array}[t]{lcl}
  w_0 \, m &=& \bot \\
  w_{i+1} \, m &=& \mathbf{if}\;(\evalexpr{e}\,m=\true)
   \;\mathbf{then}\;w_i^\star\,(\sem{c}\,m)\;\mathbf{else}\;\unit\,m
   \\[1ex]
\end{array}
\end{displaymath}
\caption{\pWHILE semantics}
\label{fig:semantics}
\end{figure*}

\subsection{Target Language}
To define the target language of our transformation, we remove
probabilistic assignments and add an $\kwdassert$
instruction, giving the following grammar:
\begin{displaymath}
\begin{array}{r@{\ }l@{\quad}l}
\Cmd ::= & \Skip \\
     \mid& \Seq{\Cmd}{\Cmd}        & \mbox{sequencing}\\
     \mid& \Ass{\Var}{\Expr}        & \mbox{deterministic assignment}\\
     \mid& \Assert{\varphi}        & \mbox{assert}\\
     % \mid& \Assume{\varphi}        & \mbox{assume}\\
     % \mid& \Havoc{x}        & \mbox{havoc}\\
     \mid & \Ass{(\Var,\Var)}{\PLap_\epsilon(\Expr,\Expr)}
     &\mbox{Laplace invocation}\\
     \mid & \Ass{(\Var,\Var)}{\PExp_\epsilon(\Expr,\Expr,\Expr,\Expr)}
     &\mbox{Exponential invocation}\\
     \mid& \Cond{\Expr}{\Cmd}{\Cmd} & \mbox{conditional}\\
     \mid& \While{\Expr}{\Cmd}      & \mbox{while loop}\\
     \mid& \Return~\Expr      & \mbox{return expression}\\
\end{array}
\end{displaymath}
The semantics of this non-deterministic target language is defined in
Figure \ref{fig:target} as a
function from a memory to a set of memories. 
The $\Assert{\varphi}$
statement checks at runtime whether the predicate $\varphi$ is valid, and stops
the execution if not.
In order to distinguish
the failure of $\kwdassert$ statements from non-terminating while
loops, we lift the domain $\mathcal{P}(\Mem)$ with a $\bot$ element:
where $\bigcup_{m\in\bot}f\,m$ is defined as $\bot$ for any $f$.
We defer the presentation of the abstract procedures
$\PLap$ and $\PExp$ until the definition of the self-product construction in
\S\ref{subsec:building}.

\begin{figure*}
\begin{displaymath}
\begin{array}{l@{\;\,}c@{\;\,}l}
\\[-1.2ex]
\sem{\Skip}\; m         &=& \{m\} \\[1ex]
\sem{\Seq{c_1}{c_2}}\; m  &=& \bigcup_{m'\in\sem{c_1}\,m}\sem{c_2}\,m \\[1ex]
\sem{\Ass{x}{e}}\; m    &=& \charfun_{m\subst{x}{\evalexpr{e}\,m}}
\\[1ex]
\sem{\Assert{\varphi}}\,m &=& \mathbf{if}\;(\evalexpr{\varphi}\,m)
  ~\mathbf{then}~\{m\}
  ~\mathbf{else}~\bot\\[1ex]
\sem{\Ass{(x_1,x_2)}{\PLap_\epsilon(e_1,e_2)}}\,m &=&
  \bigcup_v m\subst{x_1}{v}\subst{x_2}{v}\subst{\alphav}{\alphav+|e_1-e_2|\epsilon}
\\[1ex]
\sem{\Ass{(x_1,x_2)}{\PExp_\epsilon(s_1,e_1,s_2,e_2)}}\,m &=&
  \begin{array}[t]{l}
    \mathbf{if}\;(\evalexpr{s_1=s_2}\,m=\true)~\mathbf{then} \\[.5ex]
    \quad \bigcup_v
    m\subst{x_1}{v}\subst{x_2}{v}\subst{\alphav}{\alphav+|e_1-e_2|\epsilon} \\
    \mathbf{else}~ \bot
  \end{array}
  \\[6.5ex]
\sem{\Cond{e}{c_1}{c_2}}\ m &=& \mathbf{if}\;(\evalexpr{e}\,m=\true)
  ~\mathbf{then}~(\sem{c_1}\,m)
  ~\mathbf{else}~(\sem{c_2}\,m)\\[1ex]
\sem{\While{e}{c}}\ m &=&
\bigsqcup \: \hat{w}_i \, m\\[-6pt]
\end{array}
\end{displaymath}
\begin{displaymath}
\qquad \mbox{where }  \; 
\begin{array}[t]{lcl}
  \hat{w}_0 \, m &=& \emptyset \\
  \hat{w}_{i+1} \, m &=& \mathbf{if}\;(\evalexpr{e}\,m=\true)
   \;\mathbf{then}\;\bigcup_{m'\in(\sem{c}\,m)} \hat{w}_i\,m'\;\mathbf{else}\;\{m\}
   \\[1ex]
\end{array}
\end{displaymath}
\caption{Semantics of the target language}
\label{fig:target}
\end{figure*}

The enforcement of safety properties over this target language is
formalized by a standard Hoare logic, with judgments of the form
\[
  \HL{\Pre}{c}{\Post}.
\]
Here the pre- and post-conditions $\Pre$ and $\Post$ are standard
\emph{unary} predicates over memories. Hoare logic judgments can be derived
using the rules in Fig.~\ref{fig:hoare-logic}; by the standard soundness of
Hoare logic, the derivability of a judgment $\HL{\Pre}{c}{\Post}$ entails the
correctness of $c$ with respect to its specification $\Pre,\Post$.  

\begin{figure*}
\begin{displaymath}
\begin{array}{c}
  \infrule{}{\HL{\Pre}{\Skip}{\Pre}}
\qquad
  \infrule{
  }{
    \HL{\Post\subst{x}{e}}{\Ass{x}{e}}{\Post}
  }
\qquad
  \infrule{
  }{
    \HL{\Post\land\varphi}{\Assert{\varphi}}{\Post}
  }
% \qquad
%   \infrule{
%   }{
%     \HL{\Pre}{\Assume{\varphi}}{\varphi\land\Pre}
%   }
\\[3ex]
  \infrule{
    \HL{\Pre}{c_1}{\varphi} \qquad 
    \HL{\varphi}{c_2}{\Post}
  }{
    \HL{\Pre}{\Seq{c_1}{c_2}}{\Post}
  }
\qquad
  \infrule{
    \HL{\Pre\land b}{c_1}{\Post}
    \qquad
    \HL{\Pre\land \neg b}{c_2}{\Post}
  }{
    \HL{\Pre}{\Cond{b}{c_1}{c_2}}{\Post}
  }
\\[3ex]
  \infrule{
    \Pre \land v\leq 0 \Rightarrow \neg b \qquad
    \HL{\Pre \land b \land v = k}{c}{\Pre \land v < k}
  }{
    \HL{\Pre}{\While{b}{c}}{\Pre \land \neg b}
  }
\qquad
\infrule{
  \HL{\Pre'}{c}{\Post'} \qquad \Pre\Rightarrow\Pre' \qquad \Post'\Rightarrow\Post
}{
  \HL{\Pre}{c}{\Post}
}
\end{array}
\end{displaymath}
\caption{Hoare logic for non-probabilistic programs}
\label{fig:hoare-logic}
\end{figure*}

\subsection{Product Construction}
Before we define the product transformation from \pWHILE to our target language,
let us first review some preliminaries about product programs.

Product programs have been successfully used to verify $2$-safety
properties like information-flow, program equivalence, and program
robustness. As mentioned in in the introduction, a {\em synchronized} product
program can be used to simulate two runs of the same program, interleaving the
two executions and often simplifying the verification effort.  This technique,
however, has been mostly used in the verification of non-probabilistic programs. 
In the rest of this section we provide a brief introduction to
relational verification by product construction and then extend the approach to
handle quantitative reasoning over probabilistic programs.

A simple but necessary concept for the product construction is {\em
  memory separability}: we say that two programs are {\em separable}
if they manipulate disjoint sets of program variables.
In order to achieve separability in the construction of the product of
a program with itself, program variables are duplicated and marked with a left
($-_{1}$) or right ($-_{2}$) tag. For any program expression $e$ or predicate
$\varphi$, we let $e_i$ and $\varphi_i$ stand for the result of renaming every
program variable with the tag ${-}_i$.

Similarly, we say that two memories are \emph{disjoint} when their
domains (the sets of variables on which they are defined) are
disjoint. Given two disjoint memories $m_1$ and $m_2$, we
can build a memory $m=m_1\oplus m_2$ representing their (disjoint) union.
In the following, we exploit separability and use predicates to
represent \emph{binary relations} over disjoint memories $m_1$ and
$m_2$. We will suggestively write $m_1\, \Phi\, m_2$ to denote the
unary predicate $\Phi (m_1\oplus m_2)$ over the combined memory $m_1\oplus m_2$.

Given two deterministic programs $c_1$ and $c_2$, a general {\em
  product program} $c_1\times c_2$ is a syntactic construction that
merges the executions of $c_1$ and $c_2$; this construction is
required to correctly represent every pair of executions of $c_1$ and
$c_2$. Traditional program verification techniques can then be used to enforce a
relational property over $c_1$ and $c_2$.

In self-composition~\cite{BartheDR04,DarvasHS05}, the product construction $c_1\times c_2$ is
defined simply by the sequential composition $\Seq{c_1}{c_2}$. 
%% Then,
%% for instance, the verification of non-interference property with
%% respect to a set of variables $\vec{x}$ can be verified by proving a
%% Hoare judgment of the form
%% $$\HL{\vec{x}_1=\vec{x}_2}{\Seq{c_1}{c_2}}{\vec{x}_1=\vec{x}_2}.$$ 
An inconvenience of self-composition is that the verification of
$\Seq{c_1}{c_2}$ usually requires independent functional reasoning
over $c_1$ and $c_2$.
The synchronized product construction solves this problem by {\em
  interleaving} execution of two
runs of the same program---by placing corresponding pieces of the two
executions of a program close together, synchronized product programs
can more easily maintain inductive invariants relating the two runs.
% (see Appendix~\ref{app:sync:example})
%
Not only does synchronization reduce the verification effort, we will soon see
that synchronization is the key feature that enables our verification approach.

% Locating corresponding
% probabilistic operations close together allows both invocations to be
% replaced by ghost code to manage the privacy cost, as explained in
% more detail below.

\subsection{Building the Product} \label{subsec:building}

We embed the quantitative reasoning on probabilistic programs by
introducing the special program variables $\alphav$ and $\deltav$,
which serve to accumulate the privacy cost. For every statement $c$,
the self-product $\SProd{c}$ is formally defined by the rules shown in
Fig.~\ref{fig:sprod}. In a nutshell, the deterministic fragment of the
code is duplicated with appropriate variable renaming with the flags
$-_1$ and $-_2$, and the control flow is \emph{fully synchronized},
i.e., the two executions of the same program must take all the same
branches---we use the $\kwdassert$ statements to enforce this
property.

Moreover, for the self-product of a program $c$ to correctly represent two
executions of itself, we require that loop guards do not depend on
probabilistically sampled values; we assume in the remainder of this work that
the programs under verification satisfy this condition. Additionally, the
soundness of the method relies on the fact that all verified programs are
terminating, which is enforced by the Hoare logic rules in
Fig.~\ref{fig:hoare-logic}.

The probabilistic constructions are mapped to invocations to the
abstract procedures $\PLap$ and $\PExp$. The semantics of these
procedures is non-deterministic, in order to simulate sampling from a
probability distribution.  We axiomatize these abstract procedures with Hoare
specifications: Figure~\ref{fig:hoareLapExp} gives the new specifications.
Notice that both abstract procedures have a side effect: they increment the
privacy budget variable $\alphav$. In Section~\ref{sec:ptr}, we introduce a
alternative specification for $\PLap$ that also increments the budget variable
$\deltav$.

\begin{figure*}
\begin{displaymath}
  \begin{array}{c}
    \infrule{ 
    }{
      \HL{\alphav=\epsilon_0\land\deltav=\delta_0}
         {\Ass{(x_1,x_2)}{\PLap_\epsilon(e_1,e_2)}}
         {x_1=x_2 \land \alphav=\epsilon_0+|e_1-e_2|\epsilon\land\deltav=\delta_0}
    }
\\[3ex]
    \infrule{ 
    }{
      \HL{s_1=s_2\land\alphav=\epsilon_0\land\deltav=\delta_0}
         {\Ass{(x_1,x_2)}{\PExp_\epsilon(s_1,e_1,s_2,e_2)}}
         {x_1=x_2 \land \alphav=\epsilon_0+\epsilon\, \max_{r}|s_1(x_1, r) - s_2( x_2, r)| }
    }
  \end{array}
\end{displaymath}
%\hfill
%where $c = \max_{x_1,x_2,r} \frac{|s_1\,x_1\, r - s_2\, x_2\, r|}{|x_1-x_2|}$
\caption{Hoare specification for $\PLap$ and $\PExp$}
\label{fig:hoareLapExp}
\end{figure*}

\begin{figure}
\begin{displaymath}
\begin{array}{l@{=}l}
\SProd{\Skip} & 
  \begin{array}{l}
    \Skip
  \end{array}
\\[.5ex]
\SProd{\Seq{c_1}{c_2}} & 
  \begin{array}{l}
    \Seq{\SProd{c_1}}{\SProd{c_2}}
  \end{array}
\\[.5ex]
\SProd{\Ass{x}{e}} & 
  \begin{array}{l}
    \Seq{\Ass{x_1}{e_1}}{\Ass{x_2}{e_2}}
  \end{array}
\\[.5ex]
\SProd{\Rand{x}{\mathsf{Lap_\epsilon}(e)}} &
  \begin{array}{l}
    \Ass{(x_1,x_2)}{\PLap(e_1,e_2)}
  \end{array}
  % \begin{array}[t]{l}
  %   \Havoc{x_1};\\
  %   \Ass{x_2}{x_1};\\
  %   \Ass{\alphav}{\alphav+|e_1-e_2|\epsilon}
  % \end{array}
\\[.5ex]
  % \Call{(x_1,x_2,\alpha)}{\mathsf{PLap}}{e_1,e_2,\alpha}\\
% \SProd{\Rand{x}{\mathsf{Gau_{\epsilon,\delta}}(e)}} &
%   \begin{array}[t]{l}
%     \Havoc{x_1};\\
%     \Ass{x_2}{x_1};\\
%     \Ass{\alphav}{\alphav+|e_1-e_2|\epsilon};\\
%     \Ass{\deltav}{\deltav+\delta\big ( \frac{\exp{|e_1-e_2|\epsilon}-1}{\exp{\epsilon}-1} \big )}
%   \end{array}
% \\[.5ex]
\SProd{\Rand{x}{\mathsf{Exp_{\epsilon}}(s,e)}} &
  \begin{array}{l}
    \Ass{(x_1,x_2)}{\PExp(s_1,e_1,s_2,e_2)} 
  \end{array}
% \begin{array}[t]{l}
%     \Assert{s_1=s_2}; \\
%     \Havoc{x_1};\\
%     \Ass{x_2}{x_1};\\
%     \Ass{\alphav}{\alphav+  |e_1 -e_2| \,  c\, \epsilon}; \\
% \mbox{where}~ c= \max_{x_1,x_2,r} \frac{|s_1\,x_1\, r - s_2\, x_2\, r|}{|x_1-x_2|}
%   \end{array} 
\\[.5ex]
\SProd{\Cond{b}{c}{d}} &
  \begin{array}[t]{l}
    \Assert{b_1=b_2};\\
    \Cond{b_1}{\SProd{c}}{\SProd{d}}
  \end{array}
\\[.5ex]
\SProd{\While{b}{c}} &
  \begin{array}[t]{l}
    \Assert{b_1=b_2};\\
    \While{b_1}{}\\
    \quad \Seq{\SProd{c}}{\Assert{b_1=b_2}}
  \end{array}
\end{array}
\end{displaymath}
\caption{Self-product construction}
\label{fig:sprod}
\end{figure}

% We say that a predicate $\Phi$ \emph{implies an adjacency} relation on
% memories if there is a metric on memories $d$ such that for memories
% $m_1$ and $m_2$, we have $m_1\,\Phi\,m_2 \Rightarrow d(m_1,m_2)\leq 1$.

\subsection{An alternative characterization of privacy}

For the proof of soundness, we will use an alternative characterization
of $(\epsilon,\delta)$-differential privacy based on the notion of
$\epsilon$-distance. This notion is adapted from the asymmetric notion
of distance used by Barthe et al.~\cite{BartheKOZ13}.
% \mg{I think it is ok to introduce the $\epsilon$-distance here but we are using
%   distribution notation that is introduced only later. Perhaps the best is to
%   postpone or to use a more general notation.}
% \jh{Decided to move to appendix.}
\begin{definition}[$\epsilon$-distance]\label{def:alpha-dist}
  The $\epsilon$-distance $\Delta_\epsilon$ is defined as
  $$ 
  \gdist{\epsilon}{\mu_1}{\mu_2} \eqdef 
  \max_{S \subseteq A}\,(\mu_1\,S - \exp{\epsilon}\,\mu_2\,S),
  $$
  where $\mu\,S \eqdef \sum_{a \in S} \mu\,a$. We define max over an empty set
  to be $0$, so $\Delta_\epsilon(\mu_1, \mu_2) \geq 0$.
\end{definition}

By the definition of $\epsilon$-distance, a
probabilistic program $c$ is $(\epsilon, \delta)$-differentially
private with respect to $\epsilon> 0$, $\delta\geq 0$, and a relation
$\Phi$ on the initial memories of $c$ if for every two memories $m_1$
and $m_2$ related by $\Phi$, we have 
$$\gdist{\epsilon}{{\sem{c}~m_1}}{{\sem{c}~m_2}}\leq \delta.$$

The proof of our main theorem relies on a {\em lifting} operator that
turns a relation on memories into a relation on {\em distributions} over memory. 
Given a relation on memories $\Post$, and real values
$\epsilon,\delta$ we define the lifted relation on memory
distributions $\glift{\Post}{}{}{\epsilon}{\delta}$ as follows.
\begin{definition}
  For all memory distributions $\mu_1,\mu_2$,
  $\glift{\Post}{\mu_1}{\mu_2}{\epsilon}{\delta}$ if there exists $\mu$ such
  that:
  \begin{enumerate}
  \item $\pi_i\,\mu\leq\mu_i$,
  \item $\forall m_1, m_2.\, \mu (m_1 \oplus m_2) \Rightarrow m_1\, \Post\,m_2$,
    and
  \item $\Delta_\epsilon(\mu_i,\pi_i\,\mu)\leq\delta$,
  \end{enumerate}
  where 
  \begin{itemize}
    \item $(\pi_1\,\mu)\,m_1=\sum_{m_2\in\Mem}\mu\,(m_1,m_2)$, and 
    \item $(\pi_2\,\mu)\,m_2=\sum_{m_1\in\Mem}\mu\,(m_1,m_2)$.
  \end{itemize}
\end{definition}
Notice that $\epsilon$-distance between distributions is closely related to the
lifting of the equality relation, i.e.,\,
\begin{equation}
\glift{=}{\mu_1}{\mu_2}{\epsilon}{\delta}
\quad
\Longleftrightarrow
\quad
\gdist{\epsilon}{\mu_1}{\mu_2}\leq \delta.
\end{equation}
Note that the second equation is precisely the condition on output distributions
needed for $(\epsilon, \delta)$-differential privacy.
% \mg{Is this the main observation, right? Maybe we can state it as a
%   remark that we can refer later when we introduce the
%   self-product. For the moment I just put an equation.}
% \jh{Will move to appendix.}

\subsection{Soundness of the self-product technique}
We can now state the soundness theorem for our approach. Recall that we
consider only programs with a single return statement; we will label this
returned value $\outv_1$ and $\outv_2$ in the first and second runs,
respectively.

\begin{theorem}
\label{thm:hldiffpriv}
If the following Hoare judgment is valid
\begin{displaymath}
  \HL{\Pre\land \alphav\!=\!0 \land \deltav\!=\!0}
  {\SProd{c}}
  {\outv_1\!=\!\outv_2\land\alphav\!\leq\!\epsilon\land\deltav\!\leq\!\delta}
\end{displaymath}
then $c$ satisfies $(\epsilon,\delta)$-differential privacy.
\end{theorem}
The proof of Theorem~\ref{thm:hldiffpriv} follows from the next lemma.
\begin{lemma}
Let $\Post$ be a relation on memories, and suppose
  \[
    \HL{\Pre\land \alphav=0 \land \deltav=0}
    {\SProd{c}}
    {\Post\land\alphav\leq\epsilon\land\deltav\leq\delta}.
  \]
  Then, for all memories $m_1, m_2$ such that $m_1\,\Pre\,m_2$, we have
  \[
    \glift{\Post}{(\sem{c}\,m_1)}{(\sem{c}\,m_2)}{\epsilon} {\delta}.
  \]
\end{lemma}
  The lemma is proved by structural induction on $c$; we provide
  technical details in the full version of the paper.

\section{Comparison with {\sf apRHL}} \label{sec:aprhl}

Now that we have defined our transformation, we compare our approach to a custom
logic for verifying privacy. {\sf apRHL}~\cite{BartheKOZ13} is a quantitative,
probabilistic and relational program logic for reasoning about differential
privacy, with judgments of the form\footnote{The original {\sf apRHL} rules are
  based on a multiplicative privacy budget. We adapt the rules to an
  additive privacy parameter for consistency with the rest of the
  article.}
\[
  \AEquiv{\Pre}{c_1}{c_2}{\Post}{\alpha}{\delta} ,
\]
where $c_1$ and $c_2$ are probabilistic programs, $\Pre$ and $\Post$
are memory relations, and $\epsilon,\delta$ are real values. The main
result of {\sf apRHL} states that if
$\AEquiv{\Pre}{c_1}{c_2}{\outv_1=\outv_2}{\epsilon}{\delta}$ is
derivable, where $c_1$ and $c_2$ are the result of renaming variables
in $c$ to make them separable, then $c$ is
$(\epsilon,\delta)$-differentially private with respect to the relation $\Pre$
on initial memories.

\begin{figure*}[t]
$$
\begin{array}{c@{}}
\infrule
 { }
 {
   \AEquiv{\Post\subst{x_1}{e_1}\subst{x_2}{e_2}}
     {\Ass{x_1}{e_1}}{\Ass{x_2}{e_2}}{\Post}{0}{0}
   }
[\mathsf{assn}]
\\[4ex]
\infrule {
  }
  {\AEquiv{\true}
   {\Rand{y_1}{\mathsf{Lap}_\epsilon(e_1)}}{\Rand{y_2}{\mathsf{Lap}_\epsilon(e_2)}}{y_1=y_2}
   {|e_1 -e_2|\epsilon}{0}}
[\mathsf{lap}]
\\[4ex]
% \infrule {
%   }
%   {\AEquiv{\true}
%    {\Rand{y_1}{\mathsf{Gau}_{\epsilon,\delta}(e_1)}}{\Rand{y_2}{\mathsf{Gau}_{\epsilon,\delta}(e_2)}}
%    {y_1\sidel=y_2\sider}
%    {|e_1-e_2|\epsilon}{\delta}}
% [\mathsf{gau}]
% \\[4ex]
\infrule {}
  {\AEquiv{s_1= s_2}
   {\Rand{y_1}{\mathsf{Exp}_{\epsilon}(s_1,e_1)}}{\Rand{y_2}{\mathsf{Exp}_{\epsilon,s}(s_2,e_2)}}
   {y_1=y_2}
   {\epsilon\,\max_{r} {|s_1(x_1, r) - s_2( x_2, r)|} 
    }{0}}
[\mathsf{exp}]

\\[4ex]
%
% \infrule
%  {\forall m_1\ m_2.\ m_1~\Pre~m_2 \implies
% \gdist{\alpha}{\sem{\mu_1}\ m_1}{\sem{\mu_2}\ m_2} \leq \delta}
%  {\AEquiv{\Pre}
%     {\Rand{x_1}{\mu_1}}{\Rand{x_2}{\mu_2}}{x_1\sidel = x_2\sider}{\alpha}{\delta}}
% [\mathsf{rand}]\\[4ex]
%
\infrule{}{\AEquiv{\Pre}{\Skip}{\Skip}{\Pre}{0}{0}}[\mathsf{skip}]
%
% \quad
% %
% \infrule{\Pre \implies b\sidel \equiv b'\sider}
%         {\AEquiv{\Pre}{\Assert{b}}{\Assert{b'}}{\Pre\land b\sidel}{0}{0}}
% [\mathsf{assert}]
\\[4ex]
\infrule
{\AEquiv{\Pre \land b_1}{c_1}{c_2}{\Post}{\epsilon}{\delta} 
\quad 
 \AEquiv{\Pre \land \lnot b_1}{d_1}{d_2}{\Post}{\epsilon}{\delta} 
}
  {\AEquiv{\Pre\land b_1 = b_2}{\Cond{b_1}{c_1}{d_1}}{\Cond{b_2}{c_2}{d_2}}{\Post}{\epsilon}{\delta}}
[\mathsf{cond}]\\[4ex]
\infrule
{
 \begin{array}{l}
 \AEquiv{\Inv \land b_1 \land k = e}
         {c_1}{c_2}
         {\Inv \land k < e}
         {\epsilon}{\delta} \\
 \Inv \land n \leq e \implies \neg b_1 \qquad
 \Inv \implies b_1 = b_2
 \end{array}
}{
 \AEquiv{\Inv \land 0 \leq e}
         {\While{b_1}{c_1}}{\While{b_2}{c_2}}
         {\Inv \land \neg b_1}
         {n\epsilon}{n \delta}
}
[\mathsf{while}] \\[4ex]

\infrule
{\AEquiv{\Pre}{c_1}{c_2}{\Post'}{\epsilon}{\delta} 
  \quad \AEquiv{\Post'}{c_1'}{c_2'}{\Post}{\epsilon'}{\delta'} 
}
{\AEquiv{\Pre}{c_1;c_1'}{c_2;c_2'}{\Post}
  {\epsilon + \epsilon'}{\delta + \delta'}}
[\mathsf{seq}]\\[4ex]

% \infrule 
%   {\AEquiv{\Pre \land \Theta}{c_1}{c_2}{\Post}{\alpha}{\delta} \qquad 
%   \AEquiv{\Pre \land \lnot\Theta}{c_1}{c_2}{\Post}{\alpha}{\delta}}
%   {\AEquiv{\Pre}{c_1}{c_2}{\Post}{\alpha}{\delta}}
% [\mathsf{case}] \\[4ex]

% check or the alpha/delta ....
% \infrule{\AEquiv{\Pre}{c_1}{c_2}{\Post}{\alpha}{\delta} \qquad
% \AEquiv{\Pre'}{c_2}{c_3}{\Post'}{\alpha'}{\delta'}} 
%   {\AEquiv{\Pre\circ \Pre'}{c_1}{c_3}{\Post\circ\Post'}{\alpha \alpha'}{\max (\delta + \alpha\ \delta', 
% \delta' + \alpha'\ \delta )}}
% [\mathsf{comp}] \\[4ex]

\infrule {
  \AEquiv{\Pre'}{c_1}{c_2}{\Post'}{\epsilon'}{\delta'} \qquad
  \Pre \Rightarrow \Pre'  \qquad \Post' \Rightarrow \Post \qquad
  \epsilon' \leq \epsilon \qquad \delta' \leq \delta
  }
 {
   \AEquiv{\Pre}{c_1}{c_2}{\Post}{\epsilon}{\delta}
 }
[\mathsf{weak}] 

% \quad

% \infrule{\AEquiv{\Pre^{-1}}{c_2}{c_1}{\Post^{-1}}{\alpha}{\delta}} 
%   {\AEquiv{\Pre}{c_1}{c_2}{\Post}{\alpha}{\delta}}
% [\mathsf{transp}] 

% \\[4ex]

% \infrule{\AEquiv{\Pre}{c_1}{c_2}{\Post}{\alpha}{\delta}\qquad 
% \forall m_1\ m_2.\ m_1\ \Theta\ m_2 \implies \mathsf{\range} \Theta\
% (\sem{c_1}~m_1\times \sem{c_2}~m_2)}
%          {\AEquiv{\Pre\land \Theta}{c_1}{c_2}{\Post\land\Theta}{\alpha}{\delta}}
%          [\mathsf{frame}]
\end{array}
$$
\hrule
\caption{Core proof rules of the approximate relational Hoare logic}
\label{fig:aprhl}
\end{figure*}

The original presentation of the {\sf apRHL} logic~\cite{BartheKOZ13}
is organized in three sets of rules: the first set includes a set of
core rules, the second set includes a generalized rule for loops (see
Fig.~\ref{fig:looprule}), and the third set includes rules for
mechanisms such as the Laplace and Exponential Mechanism. We refer to
the fragment consisting of the first and third set of rules as {\em
  core {\sf apRHL}}; its rules are displayed in
Fig.~\ref{fig:aprhl}. Note that the, in contrast
with~\cite{BartheKOZ13}, the rule for sequential composition does not
have any side condition; this is due to the fact that the rule for
random assignments in~\cite{BartheKOZ13} allows sampling from strict
sub-distributions, whereas we only allow sampling using the Laplace
and Exponential mechanisms.

The following lemma shows that our approach subsumes core {\sf apRHL}, in the
sense that every probabilistic program $c$ verified
$(\epsilon,\delta)$-differentially private using core {\sf apRHL} can be
verified using our self-product technique.

\begin{lemma}
  For every probabilistic program $c$, memory relations $\Pre,\Post$
  and real expressions $\epsilon,\delta$ such that the following core
  {\sf apRHL} judgment is derivable
\[
  \AEquiv{\Pre}{c}{c}{\Post}{\epsilon}{\delta}
\]
we have
\[
  \HL{\Pre}{\SProd{c}}{\Post\land\alphav\leq\epsilon\land\deltav\leq\delta}.
\]
\end{lemma}
The proof of this result is straightforward, by induction on the
derivation of the apRHL judgement.
%

% Interestingly, the {\sf apRHL} rules shown in Fig.~\ref{fig:aprhl} fail to
% verify the \emph{smart sum} algorithm we describe in \S\ref{sec:sums}.
% To deal with this example, Barthe et al.~\cite{BartheKOZ13} introduce
% a generalized while rule. In contrast, the verification of
% \emph{smart-sum} using self-products and Hoare-logic does not require
% any special verification rule, as shown in \S\ref{sec:sums}. The
% additional generality of our method stems from the fact that privacy
% consumption in {\sf apRHL} is tracked by an accumulator which is part of the
% judgment but independent of the pre-condition and the initial
% memory. Using self-products, the reasoning over the privacy budget
% accumulator is carried out in the Hoare specification and consequently
% inherits the expressivity of the Hoare logic.

The embedding is more expressive than core {\sf apRHL} in its
treatment of loops. This is because privacy consumption in core {\sf
  apRHL} is tracked by an accumulator which is part of the judgment
itself, independent of the pre-condition and the initial memory.
Using self-products, reasoning about the privacy budget is carried out
in the Hoare specification and consequently inherits the full
expressivity of the Hoare logic. On the other hand, we have not been
able to capture the generalized rule for loops from {\sf apRHL}, which
is given in Fig.~\ref{fig:looprule}, with self-products. In the
following section, we provide a more detailed comparison with {\sf
  apRHL} based on examples.

We conclude with a broader perspective on the two formalisms. The
primary goal of our approach is to strike a good balance between
expressivity and simplicity, including for the latter ease of use and
ease of implementation. In contrast to {\sf apRHL}, which requires a
relational verification infrastructure, our approach reuses a very
standard verification technology, namely Hoare logic, and can be
directly implemented by defining the appropriate program
transformation, and using off-the-shelf tools for Hoare logic or even
invariant generation.  We believe this latter approach is simpler to
deploy for programming languages for which verification environments
based on Hoare logic are already available.

 \begin{figure*}
 \hrule
 $$
 \infrule
 {\begin{array}{@{}l@{}}
  \Inv \implies 
  b_1\sidel \equiv b_2\sider \land P\sidel \equiv P\sider \land i\sidel = i\sider
  \qquad
    \Inv \land n \leq i\sidel \implies \lnot b_1\sidel \\
    \AEquiv{\Inv \land (b_1 \land i = j \land \lnot P)\sidel}
    {c_1;\, \Assert{\lnot P}}
    {c_2;\, \Assert{\lnot P}}
    {\Inv \land i\sidel = j \!+\! 1}
    {\epsilon_j}{0} \\
 \AEquiv{\Inv \land (b_1 \land i = j \land \lnot P)\sidel}
         {c_1;\, \Assert{P}}
         {c_2;\, \Assert{P}}
         {\Inv \land i\sidel = j \!+\! 1}
         {\epsilon}{0} \\
 \AEquiv{\Inv \land (b_1 \land i = j \land P)\sidel}
         {c_1}
         {c_2}
         {\Inv \land (i = j \!+\! 1 \land P)\sidel}
         {0}{0}
 \end{array}}
 {\AEquiv{\Inv \land i\sidel = 0}
         {\While{b_1}{c_1}}
         {\While{b_2}{c_2}}
         {\Inv \land \lnot b_1\sidel}
         {\epsilon+ \sum_{i = 0}^{n-1} \epsilon_i}{0}}
 [\mathsf{gwhile}]
 $$
 
 \hrule
 \caption{Generalized rule for loops}
 \label{fig:looprule}
 \end{figure*}

%% Another instance of an algorithm that cannot be verified in core {\sf apRHL}
%% is the \emph{minimum vertex cover} algorithm developed by Gupta et
%% al.~\cite{GLMRT10}. The algorithm can proved differentially private in
%% an ad hoc extension of {\sf apRHL}. One can extend self-products to consider
%% the minimum vertex cover algorithm, and prove the pre-condition of
%% Theorem~\ref{thm:hldiffpriv}; however, extending the proof of
%% Theorem~\ref{thm:hldiffpriv} to account for this example is
%% problematic. \S\ref{sec:vertexcover} discusses this example in more 
%% detail.

\section{Examples} \label{sec:examples}
In this section, we apply our method to four examples. The first
example (smart sum) is an algorithm for computing statistics; it
involves intricate applications of the composition theorem, and is
thus an interesting test case. The second example (Iterative Database
Construction, or more precisely the Multiplicative Weights Exponential
Mechanism) is an algorithm that computes a synthetic database; it
combines the Laplace and the Exponential mechanisms, and has not been
verified in earlier work using relational logic. The third example
(Propose-Test-Release) is an algorithm that only achieves approximate
differential privacy (i.e., $(\epsilon,\delta)$-differential privacy
with $\delta>0$) using both the privacy and accuracy properties of the
Laplace distribution. To best of our knowledge, we provide the first machine-checked
proof of this mechanism. Finally, our last example (vertex cover) is
an algorithm that achieves differential privacy by carefully adding
noise to sampled values; this example can only be verified partially
using our method, and illustrates the differences with {\sf apRHL}.

\subsection{Smart sum}
\label{sec:sums}
In this example, a database $\mathsf{db}$ is a list of real numbers
$[r_1,\ldots,r_T]$ and we consider two databases {\em adjacent} if
they are the same length $T$, at most one entry differs between the
two databases, and that entry differs by at most $1$.

Suppose we want to release private sums of the first $i$ entries, simultaneously
for every $i \in [1 \dots T]$: that is, given $[r_1,r_2,r_3,r_4,\ldots,r_T]$ we
want to privately release
\[
  \left[r_1,\sum_{i=1}^2r_i,
    \sum_{i=1}^3r_i,\sum_{i=1}^4r_i,\ldots,\sum_{i=1}^Tr_i\right].
\]

An interesting sophisticated differentially private algorithm for this problem
is the {\em two-level counter} from Chan, et al.~\cite{chan-counter}; we call
this algorithm $\mathsf{smartsum}$.

At a high level, this algorithm groups the input list into blocks of length $q$,
and adds Laplace noise to the  sum for each block.  More concretely, to compute
a running sum from $1$ to $t$ with $t$ a multiple of $q$, we simply add together
the first $t/q$ block sums. If $t$ is not a multiple of $q$, say $t = qs + r$
with $r < q$, we take the first $s$ block sums and add a noised version of each
of the $r$ remaining elements. 

For an example, suppose we take $q=3$ and $T$ is a multiple of $3$. For
brevity, let us use the notation $\mathsf{L}(r)$ to describe the result of the
application of Laplace, for a fixed value $\epsilon$ to $r$. Then, the output of
$\mathsf{smartsum}$ is
\begin{multline*}
  \left[
    \mathsf{L}\left(r_1\right),\mathsf{L}\left(r_1\right)+\mathsf{L}\left(r_2\right),
    \mathsf{L}\left(\sum_{i=1}^3r_i\right),\right.
  \\
  \left. \mathsf{L}\left(\sum_{i=1}^3r_i\right)+\mathsf{L}\left(r_4\right),
    \ldots,\sum_{j = 0}^{T/3}\mathsf{L}\left(\sum_{i=1}^3r_{3j+i}\right)
  \right].
\end{multline*}

To informally argue privacy, observe that if we run the Laplace
mechanism on each individual entry, there is no privacy cost for the indices
where the adjacent databases are the same.  So, the privacy analysis for
$\mathsf{smartsum}$ is straightforward: changing an input element will change
exactly two noisy sums---the sum for the block containing $i$, and the noisy
version of $i$---and each noisy sum that can change requires $\epsilon$ privacy
budget, since we are using the Laplace mechanism with parameter $\epsilon$.
Thus, $\mathsf{smartsum}$ is $2\epsilon$-private.

The full program, together with the transformation into a synchronized product
program, is presented in Fig.~\ref{fig:smart}.  The formal verification of the
$2\epsilon$-differential privacy follows the argument above.  The pre-condition
states that the two input databases are adjacent, while  the post-condition
requires equality on the outputs and bounds the accumulated privacy budget by
$2\epsilon$.  

The interesting part for our verification is the while loop.  Indeed,
this requires a loop invariant to keep track of the privacy budget,
which depends on whether the differing entry has been processed or
not. Note that this invariant does not fit the core {\sf apRHL}
$\mathsf{while}$ rule of Fig.~\ref{fig:aprhl}: to deal with this
example, Barthe et al.~\cite{BartheKOZ13} use the generalized while
rule from Fig.~\ref{fig:looprule}. This rule is able to perform a
refined analysis depending on a predicate that is preserved across the
first iterations, until some critical iteration is reached. In
contrast, here we do not require any special verification rule: the
standard while rule from Hoare logic suffices.

More precisely, we apply the Hoare while rule with the
invariant:
\begin{displaymath}
  \begin{array}{l}
    \mathsf{adjacent}(l_1,l_2) \land 
    out_1=out_2 \land 
    next_1=next_2 \land 
    n_1=n_2 \land \\ 
  |c_1 -c_2| \leq 1 \land
  (l_1\neq l_2 \Rightarrow \alphav=0 ) \land \\
  (c_1\neq c_2 \Rightarrow l_1= l_2 \land \alphav \leq \epsilon)
  \land 
  (l_1=l_2 \rightarrow \alphav \leq 2\,\epsilon)
\end{array}
\end{displaymath}
Notice from the invariant that if the accumulators $c_1$ and $c_2$
differ we have $l_1=l_2$. This corresponds to the fact that the
differing entry has been processed and so the remaining 
database entries coincide. Also, if this is the case then the privacy
budget of $2\,\epsilon$ has been already consumed.  

The verification
of this invariant proceeds by case analysis. We have three cases: a)
the differing entry has not been processed yet and will not be
processed in the following iteration, b) the differing entry has not
been processed yet but
is going to be processed in the next iteration, and c) the differing
entry has
already been processed, in which case there is no more
privacy budget consumption.
\begin{figure}
\begin{subfigure}[b]{\columnwidth}
\begin{displaymath}
\begin{array}{l}
\Ass{next}{0};
\Ass{n}{0};
\Ass{c}{0};\\
\While{0< \nm{length} l}{
  \\
  \quad \Cond{\nm{length} l\, \nm{mod} q = 0}{
    \\
    \quad\quad \Ass{x}{\nm{Lap}_\epsilon (c + \nm{hd} l)};\\
    \quad\quad \Ass{n}{x + n};\\
    \quad\quad \Ass{next}{n};\\
    \quad\quad \Ass{c}{0};\\
    \quad\quad \Ass{out}{next :: out};\\
  }{
    \\
    \quad\quad \Ass{x}{\nm{Lap}_\epsilon (\nm{hd} l)};\\
    \quad\quad \Ass{next}{next + x};\\
    \quad\quad \Ass{c}{c + \nm{hd} l};\\
    \quad\quad \Ass{out}{next :: out};\\
  }
  \quad \Ass{l}{\nm{tl} l};\\
}
\Return~out;\\
\end{array}
\end{displaymath}
\caption{Original probabilistic algorithm}
\end{subfigure}

\begin{subfigure}[b]{\columnwidth}
\begin{displaymath}
\begin{array}{l}
\Ass{\alphav}{0};\,
\Ass{next_1}{0};\,
\Ass{next_2}{0};\\
\Ass{n_1}{0};\,
\Ass{n_2}{0};\,
\Ass{c_1}{0};\,
\Ass{c_2}{0};\\
\Assert{(0< \nm{length} l_1)\Leftrightarrow (0< \nm{length} l_2)};\\
\While{0< \nm{length} l_1}{
  \\
  \quad \Assert{(\nm{length} l_1\, \nm{mod} q = 0)\Leftrightarrow (\nm{length} l_2\, \nm{mod} q = 0)};\\
  \quad \Cond{\nm{length} l_1\, \nm{mod} q = 0}{
    \\
\quad \quad \Ass{(x_1,x_2)}{\PLap_\epsilon(c_1 + \nm{hd} l_1,c_2 + \nm{hd} l_2))}; \\    
    \quad\quad \Ass{n_1}{x_1 + n_1};\, \Ass{n_2}{x_2 + n_2};\\
    \quad\quad \Ass{next_1}{n_1};\, \Ass{next_2}{n_2};\\
    \quad\quad \Ass{c_1}{0};\, \Ass{c_2}{0};\\
    \quad\quad \Ass{out_1}{next_1 :: out_1};\, \Ass{out_2}{next_2 :: out_2};\\
  }{
    \\
\quad \quad \Ass{(x_1,x_2)}{\PLap_\epsilon(\nm{hd} l_1, \nm{hd} l_2))}; \\    
    \quad\quad \Ass{next_1}{next_1 + x_1};\,\Ass{next_2}{next_2 + x_2};\\
    \quad\quad \Ass{c_1}{c_1 + \nm{hd} l_1};\, \Ass{c_2}{c_2 + \nm{hd} l_2};\\
    \quad\quad \Ass{out_1}{next_1 :: out_1};\, \Ass{out_2}{next_2 :: out_2};\\
  }
  \quad \Ass{l_1}{\nm{tl} l_1};\, \Ass{l_2}{\nm{tl} l_2};\\
}
\Return~(out_1,out_2);\\
\end{array}
\end{displaymath}
% \begin{displaymath}
% \begin{array}{l}
% \Ass{sum_1}{0};\,\Ass{sum_2}{0};\\
% \Ass{out_1}{[\,]};\,\Ass{out_2}{[\,]};\\
% \While{0 < \nm{length}l_1}{} \\
% \quad \Ass{z_1}{\nm{geom_\epsilon} (\nm{hd} l_1)};\\
% \quad \Ass{sum_1}{sum_1+z_1};\\
% \quad \Ass{out_1}{sum_1::out_1};\\
% \quad \Ass{l_1}{\nm{tl}l_1};\\
% \While{0 < \nm{length} l_2}{} \\
% \quad \Ass{z_2}{\nm{geom_\epsilon} (\nm{hd} l_2)};\\
% \quad \Ass{sum_2}{sum_2+z_2};\\
% \quad \Ass{out_2}{sum_2::out_2};\\
% \quad \Ass{l_2}{\nm{tl} l_2};\\
% \Return (out_1, out_2);
% \end{array}
% \end{displaymath}
\caption{Synchronized non-probabilistic product}
\end{subfigure}
% \begin{subfigure}[b]{\columnwidth}
%   \jh{Should we include the synchronized product? It will be enormous.}
% % \caption{Synchronized product}
% % \label{fig:smartprod}
% \end{subfigure}
\caption{$\mathsf{smartsum}$ algorithm}
\label{fig:smart}
\end{figure}

\subsection{Multiplicative Weights Exponential Mechanism}
While answering queries on a database with the Laplace mechanism is a
simple way to guarantee privacy, the added noise quickly renders the
results useless as the number of queries grows. To handle larger
collections of queries, there has been much research on sophisticated
algorithms based on learning theory.

One such scheme is {\em Iterative Database Construction (IDC)}, due to
Gupta et al.~\cite{GRU12}. The basic idea is simple: given a database
$\hat{d}$, the algorithm gradually builds a \emph{synthetic database}
that approximates the original database. The synthetic database is
built over several rounds; after some fixed number of rounds, the
synthetic database is released and used to answer all queries.

The essence of the algorithm is the computation that it performs at each
round. Let $\mathcal{Q}$ be a collection of queries that we want to
answer and let $d^i$ be the synthetic database computed at round $i$.
During round $i+1$, the algorithm selects a query $q \in \mathcal{Q}$
with high error; that is, a query where the current
approximate database $d^i$ and the true database $\hat{d}$ give very
different answers. This selection is done in a differentially private
way. Next, the algorithm computes a noisy version $v$ of $q$ evaluated
on the true database $\hat{d}$. Again, this step must be
differentially private. Finally, $q$, $v$ and the current database
$d^i$ approximation are fed into an update algorithm, which generates
the next approximation $d^{i+1}$ of the synthetic database (hopefully
performing better on $q$).

The idea is that in many cases, this iterative procedure will provably find an
approximation with low error on {\em all} queries in $\mathcal{Q}$ in a small
number of steps. Hence, we can run IDC for a small number of steps, and release
the final database approximation as the output. Queries in
$\mathcal{Q}$ can then be evaluated on this output for an accurate
estimate of the true answer to the query.

IDC is actually a family of algorithms parameterized by an algorithm
to privately find a high-error query (called the {\em private
  distinguisher}), and the update function (called the {\em database
  update algorithm}). For concreteness, let us consider one
well-studied instantiation, the {\em Multiplicative Weights
  Exponential Mechanism} ({\sf MWEM}) algorithm originally due to
Hardt and Rothblum~\cite{HardtR10} and experimentally evaluated by
Hardt et al.~\cite{HardtLM12}.

{\sf MWEM} uses the exponential mechanism to privately select a query
with high error---the quality score of a query $q$ to be maximized is
the error of the query, i.e., the absolute difference between $q$
evaluated on the approximate database $d^i$ and $q$ evaluated on the
true database $\hat{d}$. The update function applies the
multiplicative weights update~\cite{mwsurvey} to adjust the approximation to
perform better on the mishandled query.
% this function is rather technical to
% describe here, but intuitively adjusts the approximation to perform
% better on the mishandled query.
This step is non-private: it does not
touch the private data directly. Hence, we do not concern ourselves
with the details here, and treat the update step as a black box. (The
reader can find further details in Hardt et al.~\cite{HardtLM12}.) The full
program, together with the transformation into a synchronized product
program, is presented in Fig.~\ref{fig:mwem}.

We briefly comment on the program. We let $d^i$ denote the $i$-th
iteration of the synthetic database, and $\hat{d}$ denote the true
database. Initially the synthetic database $d^0$ is set to some
default value $\mathrm{def}$. Then we define the score function $s^i$
that takes as inputs a database $D$ and a query $Q$ and returns the
error of the query $Q$ on the current approximation $d^i$ compared to $D$.
We then apply the exponential mechanism to the true database
$\hat{d}$ with the score function $s^i$, and we call the result $q^i$. We
then evaluate $q^i$ on the real database, and add Laplace noise; we
call the result $a^i$. Finally, we apply the update function to obtain
the next iteration $d^{i+1}$ of the synthetic database.  Once the
number of rounds is exhausted, we return the last computed synthetic
databases.

For the privacy proof, we assume that all queries in $\mathcal{Q}$ are
1-sensitive. Note that we run $T$ iterations of {\sf MWEM}; by the
composition theorem, it is sufficient to analyze the privacy budget
consumed by each iteration. Each iteration, we select a query with the
exponential mechanism with privacy parameter $\epsilon$, and we
estimate the true answer of this query with the Laplace mechanism,
parameter $\epsilon$. By the composition theorem
(Theorem~\ref{thm:compose}), the whole algorithm is private with
parameter $2 \cdot T \cdot \epsilon = 2T\epsilon$, as desired. The
proof can be transcripted directly into Hoare logic using
self-products; we take as pre-condition adjacency of the two databases,
and use adjacency to conclude that the sensitivity of the score
function $s_i$ is 1 at each iteration.

\begin{figure}
\begin{subfigure}{\columnwidth}
\begin{displaymath}
\begin{array}{l}
\Ass{i}{0}; \\
\Ass{d^0}{\mathrm{def}};\\
\While{i<T}{} \\
\quad \Ass{s^i}{\lambda D ~Q.~|Q(d^i)-Q(D)|}\; \\
\quad \Ass{q^i}{{\sf Exp}_{\epsilon}\,(s^i,\hat{d})};\\
\quad \Ass{a^i}{{\sf Lap}_{\epsilon}\,(q^i\,\hat{d})}; \\
\quad \Ass{d^{i+1}}{\nm{update}(d^i,a^i,q^i)}; \\
\quad \Ass{i}{i+1}; \\
\Return~d^T;
\end{array}
\end{displaymath}
\caption{Original probabilistic algorithm}
\end{subfigure}
\\[2ex]
\begin{subfigure}{\columnwidth}
\begin{displaymath}
\begin{array}{l}
\Ass{\alphav}{0};\ \Ass{i_1}{0};\ \Ass{i_2}{0};\\
\Ass{d^0_1}{\mathrm{def}};\Ass{d^0_2}{\mathrm{def}};\\
\Assert{i_1<T \Leftrightarrow i_2<T};\\
\While{i_1<T}{}\\
\quad \Ass{s^i_1}{\lambda D ~Q.~|Q(d^i_1)-Q(D)|}; \\
\quad \Ass{s^i_2}{\lambda D ~Q.~|Q(d^i_2)-Q(D)|}; \\
% \quad \Ass{q^i_1}{\mathsf{Exp}_{\epsilon}(s^i_1,\hat{d}_1)}; \\
% \quad \Ass{q^i_2}{\mathsf{Exp}_{\epsilon}(s^i_2,\hat{d}_2)}; \\
\quad
\Ass{(q^i_1,q^i_2)}{\PExp_\epsilon(s^i_1,\hat{d}_1,s^i_2,\hat{d}_2)}; \\
% \quad \Ass{a^i_1}{q^i_1(\hat{d}_1)}; \\
% \quad \Ass{a^i_2}{q^i_2(\hat{d}_2)}; \\
\quad \Ass{(a^i_1,a^i_2)}{\PLap_\epsilon(q^i_1(\hat{d}_1),q^i_2(\hat{d}_2))}; \\
\quad \Ass{d_1^{i+1}}{\nm{update}(d^i_1,a^i_1,q^i_1)}; \\
\quad \Ass{d_2^{i+1}}{\nm{update}(d^i_2,a^i_2,q^i_2)}; \\
\quad \Ass{i_1}{i_1+1}; \\
\quad \Ass{i_2}{i_2+1}; \\
\quad \Assert{i_1<T \Leftrightarrow i_2<T};\\
\Return~(d^T_1,d^T_2);
\end{array}
\end{displaymath}
\caption{Synchronized non-probabilistic product}
\end{subfigure}
\caption{{\sf MWEM} algorithm}
\label{fig:mwem}
\end{figure}

\subsection{Propose-Test-Release}
\label{sec:ptr}
The examples we have considered so far all rely on the composition
theorem.  While this is a quite powerful and useful theorem, not all algorithms
use composition.  In this section,
we consider one such example: the {\em Propose-Test-Release} (PTR)
framework \cite{DworkL09,ThakurtaS13}. PTR is also an example of
an $(\epsilon,\delta)$-differentially private mechanism for $\delta>0$.

The motivation comes from private release of statistics that are
sometimes, but not always, very sensitive. For example, suppose our
database is an ordered list of numbers  between $0$ and $1000$, and suppose we want to
release the median element of the database. This can be highly
sensitive: consider the database $[0, 0, 1000]$ with median
$0$. Adding a record $1000$ to the database would lead to a large
change in the median (now $500$, if we average the two elements
closest to the median when the database has even size).
However, many other databases have low sensitivities: for $[0, 10,
10, 1000]$, the median will remain unchanged (at $10$) no matter what
element we add or remove from the database. We may hope that we can
privately compute the median in this second case with much less noise
than needed for the first case.
More generally, the second database is quite {\em stable}---all
adjacent databases have the same median value. In contrast, the first
database is {\em instable}---adjacent databases may have wildly
different median values. With this example in mind, we now explain the
general PTR framework.

Suppose we want to privately release the result of a query $q$ evaluated on a
database $d$. We assume that databases are taken from a set
$\mathcal{D}$ and that there exists a notion of distance $\Delta$ on
$\mathcal{D}$, such that pairs of input memory related by $\Phi$ correspond to
databases at distance at most $1$ under $\Delta$.  First, we estimate the {\em
  distance to instability}---that is, the largest distance $x$ such that $q(d) =
q(d')$ for all databases $d'$ at distance $x$ or less from $d$. Since
this a $1$-sensitive function (moving to a neighboring database can
change the distance to instability by at most $1$), we can release
this distance privately using the Laplace mechanism (say, with
parameter $\epsilon$). Call the result $y$.
Now, we compare $y$ to a threshold $t$ (to be specified
later). If $y$ is less than the threshold, we output $q(d)$ with
no noise. If $y$ is greater than the threshold, we output a
default value $\perp$. The program is given in Fig.~\ref{fig:ptr}.
% The program, together with the transformation
% into a synchronized product program, is given in Fig.~\ref{fig:ptr}.

\begin{figure}
\begin{displaymath}
\begin{array}{l}
\Ass{x}{\nm{DistToInstability}(q, d)};\\
\Ass{y}{\nm{Lap}_\epsilon\ x}; \\
\mathsf{if}~(|y| >  \log(2/\delta)/(2\epsilon))\\
\quad \Return~(q\,d); \\
\mathsf{else} \\
\quad \Return~(\bot);\\
\end{array}
\end{displaymath}
\caption{PTR algorithm}
\label{fig:ptr}
\end{figure}

The privacy of the algorithm can be informally justified in two parts. First,
suppose that instead of outputting $q(d)$ or $\perp$, we simply output which
branch the program took.  This is $\epsilon$-differentially private: computing
$y$ is $\epsilon$-differentially private (via the Laplace mechanism),
and the resulting branch is a post-processing of $y$. Hence, we can assume
that the same branch is taken in both executions.

Second, we can conclude that the original program (outputting $q(d)$ or $\perp$)
is $(\epsilon,\delta)$-differentially private if for any adjacent databases $d$
and $d'$ with $q(d) \neq q(d')$, the first branch is taken with probability at
most $\delta$. By properties of the Laplace mechanism, we can set the
threshold $t$ large enough so that with probability at least $1 - \delta$, the
first branch is only taken if $x$ is strictly positive. In this case we can
conclude $q(d) = q(d')$, since $q(d)\neq q(d')$ implies that $x$ is $0$ on both
executions. So, we can safely release $q(d) = q(d')$ with no noise.  Of course,
if the second branch is taken, then it is also safe to release $\perp $ in both
runs.
%\mg{I think this crucial point needs some more work. I will do later.}

More formally, the proof of $(\epsilon,\delta)$-differential privacy
for PTR rests on two properties of the Laplace mechanism: the privacy
property captured by Theorem~\ref{thm:laplace} and the accuracy
property captured by Lemma~\ref{lem:accuracy}.
%% We modify the self-product construction of the Laplace mechanism in order to reflect both
%% these properties. The modified translation
%% for $y\leftarrow {\sf Lap}_{\epsilon}\, e$ is
%% as follows:
%% $$\begin{array}{l}
%% \Havoc{y_1};\\
%% \Ass{y_2}{y_1};\\
%% \Ass{\alphav}{\alphav+\epsilon | e_1 - e_2|};\\
%% \Condt{e_1=e_2}{\Assume{|y_1 - e_1| \leq t}; 
%% \Ass{\deltav}{\deltav+\delta}};
%% \end{array}$$
%% where $t=\log(2/\delta)/(2\epsilon)$. The soundness of the extended
%% construction is shown in the Appendix.

Fig.~\ref{fig:ptr:proof} presents the proof of PTR using
the synchronized product program---the code
is interleaved with some of the pre- and post-conditions. The proof uses the accuracy property of the Laplace
mechanism and the properties of the distance to instability that we
give as specifications in Fig.~\ref{fig:ptr:axioms}. For simplicity, we treat distance to
instability as an abstract procedure; however, it can be implemented
as a loop over all databases, in which case the specification can be
proved. The soundness of the accuracy specification for the Laplace
mechanism follows by Lemma~\ref{lem:accuracy}.

%% The self-product of the PTR algorithm under this extended translation
%% is given in  Fig.~\ref{fig:ptr}. The proof of the Hoare specification
%% \begin{displaymath}
%%   \HL{\Delta(d_1,d_2)\leq 1}
%%   {\SProd{\text{PTR}}}
%%   {\outv_1=\outv_2\land\alphav\leq\epsilon\land\deltav\leq\delta}
%% \end{displaymath}
%% uses the following specification for the distance to instability:
%% %
%% \begin{align*}
%% \vdash &\Ass{x_1}\nm{DistToInstability}(f, d_1);
%% \Ass{x_2}\nm{DistToInstability}(f, d_2): \\
%% &\Delta(d_1,d_2)\leq 1 \Longrightarrow  (f~d_1=f~d_2 \vee x_1=x_2=0)
%% \end{align*}
%% \mg{I would add here the hoare judgment for Laplace. Will do...The
%%   following comments are nice but I would not go too far with the
%%   privacy analysis. Our contribution is that we capture it not the
%%   algorithm in itself.}

\begin{figure}
\begin{displaymath}
\begin{array}{l} \\
\textcolor{DarkBlue}{\left\{ \Delta(d_1,d_2)\leq 1 \right\} }\\
\Ass{\alphav}{0};\ \\
\Ass{\deltav}{0};\\
\Ass{x_1}{\nm{DistToInstability}(q, d_1)};\\
\Ass{x_2}{\nm{DistToInstability}(q, d_2)};\\
\textcolor{DarkBlue}{\left\{ \begin{array}{l}
(q(d_1)=q(d_2)\vee x_1=x_2=0)\\ \wedge \alphav =0 \wedge \deltav =0
\end{array}\right\}} \\
\Ass{(y_1,y_2)}{\PLap_\epsilon (x_1,x_2)}; \\
\textcolor{DarkBlue}{\left\{ \begin{array}{l}
(q(d_1)=q(d_2))\vee (x_1=x_2=0\\\wedge |y_1-x_1| \leq \log(2/\delta)/(2\epsilon))\\ \wedge y_1=y_2 
\wedge \alphav \leq \epsilon \wedge \deltav \leq\delta\\
\end{array}\right\} }\\
~~~\Assert{|y_1|> \log(2/\delta)/(2\epsilon) \Leftrightarrow |y_2| >
  \log(2/\delta)/(2\epsilon)};\\
\mathsf{if}~(|y_1|> \log(2/\delta)/(2\epsilon))\\
~~~\textcolor{DarkBlue}{\left\{ \begin{array}{l}
q(d_1)=q(d_2) \land \alphav \leq \epsilon \wedge \deltav \leq\delta
\end{array}\right\}}
 \\
\quad \Return~(q(d_1),q(d_2)); \\
~~~\textcolor{DarkBlue}{\left\{ \begin{array}{l}
\mathsf{out}_1=\mathsf{out_2} \land \alphav \leq \epsilon \wedge \deltav \leq\delta
\end{array}\right\} }\\
\mathsf{else} \\
~~~\textcolor{DarkBlue}{\left\{ \begin{array}{l}
\alphav \leq \epsilon \wedge \deltav \leq\delta
\end{array}\right\}}
 \\
\quad \Return~(\bot,\bot);\\
~~~\textcolor{DarkBlue}{\left\{ \begin{array}{l}
\mathsf{out}_1=\mathsf{out_2} \land \alphav \leq \epsilon \wedge \deltav \leq\delta
\end{array}\right\}}
\end{array}
\end{displaymath}
\caption{Proof of Propose-Test-Release}\label{fig:ptr:proof}
\end{figure}
\begin{figure*}
\begin{displaymath}
\HL{x_1=x_2\land \deltav= \hat{\delta}}{\Ass{(y_1,y_2)}{\PLap_\epsilon (x_1,x_2)}}{
 y_1=y_2 \wedge |y_1 - x_1 | \leq \log(2/\delta)/(2\epsilon) \wedge \deltav = \hat{\delta} + \delta}
\end{displaymath}
\begin{displaymath}
\HL{\Delta(d_1,d_2)\leq 1}{
\Ass{x_1}\nm{DistToInstability}(q, d_1);
\Ass{x_2}\nm{DistToInstability}(q, d_2)}
{q(d_1)=q(d_2) \vee x_1=x_2=0}
\end{displaymath}
\caption{Accuracy specification for the Laplace mechanism, and
  specification for distance to instability.}\label{fig:ptr:axioms}
\end{figure*}

\subsection{Vertex cover}\label{sec:vertexcover}
A vertex cover for a graph $g=(N,E)$ is a set $S$ of nodes such that
for every edge $(t,u)\in E$, either $t\in S$ or $u\in S$. The
minimum vertex cover is the problem of finding a vertex cover of a
minimum size. Gupta et al.~\cite{GLMRT10} study the problem of
privately computing a minimum vertex cover in a setting where the
nodes of the graph are public, but its edges are private. Since a
vertex cover leaks information about vertices (for instance, any two
nodes that are not in the vertex cover are certainly not connected by
an edge), their algorithm outputs an enumeration of the nodes of the
graph, from which a vertex cover can be recomputed efficiently from
the knowledge of the set $E$. Their algorithm is challenging to verify because
rather than relying on mechanisms, it achieves privacy by sampling
according to a suitable noisy distribution $\mathsf{choose}$. The code
of the algorithm is shown in Fig.~\ref{fig:vertexcover}.

We say that two graphs $g_1$ and $g_2$ are adjacent if they differ at
most in one edge $\langle{t,u}\rangle$. By defining
$\mathsf{choose}$ as
\begin{align*}
\Pr{\Ass{v}{\mathsf{choose}_{\epsilon,n}(g)}}{v=v'} \propto \left(d_{E,V}(v') + \frac{4}{\epsilon} 
\sqrt{\frac{n}{|E|}}\right)
\end{align*} 
where $g=(E,V)$ and $n$ is a given parameter, one obtains  an
$(\epsilon,0)$-differentially private algorithm with respect to the
adjacency relation as defined above.

\begin{figure}
\begin{displaymath}
\begin{array}{l}
  \Ass{n}{|E|};\\
  \Ass{out}{[\,]};\\
%  \Ass{G_0}{G}; \\
  \While{g\neq\emptyset}{}\\
  \quad \Ass{v}{{\sf choose}_{\epsilon,n}(g)}; \\
  \quad \Ass{out}{v :: out}; \\
  \quad \Ass{g}{g\setminus \{v\}}; \\
  \Return~out;
\end{array}
\end{displaymath}
\caption{Minimum vertex cover}
\label{fig:vertexcover}
\end{figure}

In~\cite{BartheKOZ13}, Barthe \emph{et al} prove differential privacy
of vertex cover in {\sf apRHL}. The proof uses the generalized rule
for loops, a code motion rule that allows to swap independent
statements, and a rule for dealing with statements of the form
$\Rand{x}{\mu};\Assert{\phi}$. It also relies on {\sf apRHL}
specifications of {\sf choose}, that are proven correct in the {\sf
  Coq} proof assistant from the definition of  {\sf choose}.

We now consider the formal verification of the vertex cover algorithm
using self-products. We first extend the definition of self-product to
$\mathsf{choose}$. Then, there are two cases to consider: $g_2=g_1\cup
\{\langle{u,t}\rangle\}$ and $g_1=g_2\cup\{\langle{u,t}\rangle\}$.  In
the first case, we can use the first Hoare specification from
Fig.~\ref{fig:vertex:proof}. In the second case, we use the second and
third specifications from Fig.~\ref{fig:vertex:proof}. Using these
specifications, it is possible to verify that the self-product of the
vertex cover algorithm satisfies the Hoare specification of
Theorem~\ref{thm:hldiffpriv}. However, we have not yet been able to
extend the proof of Theorem~\ref{thm:hldiffpriv} to deal with the \nm{choose}
self-product.

\begin{figure*}
\begin{displaymath}
\HL{g_1 \cup\{\langle{u,t}\rangle\} =g_2 \land \alphav=\epsilon_0}{
    \Ass{(v_1,v_2)}{\mathsf{choose}^\diamond_{\epsilon,n}(g_1,g_2)}}{
     v_1=v_2 \wedge \alphav =\epsilon_0+ \epsilon/\left(2\sqrt{n}\sqrt{|g_1|}\right)
    }
\end{displaymath}
\begin{displaymath}
\HL{g_1 =g_2\cup\{\langle{u,t}\rangle\} \land \alphav=\epsilon_0}{
    \Ass{(v_1,v_2)}{\mathsf{choose}^\diamond_{\epsilon,n}(g_1,g_2)}}{
     (v\neq t\wedge v\neq u) \wedge \alphav =\epsilon_0}
\end{displaymath}
\begin{displaymath}
\HL{g_1 =g_2\cup\{\langle{u,t}\rangle\} \land \alphav=\epsilon_0}{
    \Ass{(v_1,v_2)}{\mathsf{choose}^\diamond_{\epsilon,n}(g_1,g_2)}}{
     (v=t\vee v=u) \wedge \alphav =\epsilon_0+ \epsilon/4}
\end{displaymath}

\caption{Hoare specifications for $\mathsf{choose}^\diamond$}\label{fig:vertex:proof}
\end{figure*}

\subsection{Formal verification of the examples}\label{sec:formalization}
The examples above (with the exception of vertex cover) have been
formally verified. For each example, we have built the corresponding
self-product program, and verified this result using the
non-probabilistic and non-relational Hoare logic rules available in
the {\sf EasyCrypt}~\cite{BartheDGKZ13} framework.  As described
above, we have used non-probabilistic axiomatic specifications for the
primitives. Apart from the axiomatic specification, and the code for
the program and the self-product construction, the longest Hoare logic
verification proof (for MWEM) consists of about 50 lines of code. This
demonstrates the simplicity offered by the self-product
construction. The code for these examples (and others) is available
online~\cite{code}.

\section{Related work}
Differential privacy, first proposed by Blum et al.~\cite{BDMN05} and formally
defined by Dwork et al.~\cite{DMNS06}, has been an area of intensive research in
the last decade. We have touched on a handful of private algorithms, including
algorithms for computing running sums~\cite{chan-counter,DNPR10} (part of a
broader literature on streaming privacy), answering large classes of
queries~\cite{HardtR10,HardtLM12} (part of a broader literature on
learning-theoretic approaches to data privacy), the Propose-Test-Release
framework for answering stable queries in a noiseless way~\cite{DworkL09,ThakurtaS13},
and private combinatorial optimization~\cite{GLMRT10}.  We refer readers
interested in a more comprehensive treatment to the excellent surveys by
Dwork~\cite{Dwork06,dpsurvey}.

\paragraph*{Verifying differential privacy}
Several tools have been proposed for providing formal verification of
the differential privacy guarantee; we can roughly classify them by
the verification approach they use. \pinq~\cite{mcsherry.pinq09}
provides an encapsulation for \nm{LINQ}---an SQL-like language
embedded in C\#---tracking at runtime the privacy budget consumption,
and aborting the computation when the budget is exhausted.
\nm{Airavat}~\cite{airavat} combines a similar runtime monitor with
access control in a MapReduce framework. While \pinq is restricted to
$\epsilon$-differential privacy, \nm{Airavat} can handle also
approximate differential privacy using a runtime monitor for $\delta$.

Another approach is based on linear type systems. {\sf Fuzz}~\cite{ReedP10} and
{\sf DFuzz}~\cite{GaboardiHHNP13} use a type-based approach for inferring and
checking the sensitivity of functional programs. This sensitivity analysis
combined with the use of trusted probabilistic primitives provides the
differential privacy guarantee.  Interestingly, this type-based approach can be
combined with type systems for cryptographic protocols to verify differential
privacy for distributed protocols~\cite{EignerM13}.  All these systems provide
automatic verification of differential privacy. However, they fail to verify all the
examples that we can handle, like advanced sum statistics~\cite{chan-counter} and
the Propose-Test-Release framework~\cite{DworkL09}.  Moreover, so far they can
address only pure differential privacy, where $\delta = 0$.

Tschantz, et al.~\cite{Tschantz201161} consider a verification framework for
interactive private programs, where the algorithm can receive new input and
produce multiple outputs over a series of steps.  They follow an approach
similar to ours by verifying the correct use of differentially private
primitives. However, their programs are well-modeled by probabilistic
I/O-automata, and they provide a proof technique based on probabilistic
bisimulation. Also, their method is currently limited to pure
differential privacy.

Finally, {\sf CertiPriv}~\cite{BartheKOZ13} and {\sf
  EasyCrypt}~\cite{BartheDGKZ13} use custom relational logics to
verify differential privacy. These systems are very expressive: they
supports general $(\epsilon, \delta)$-differential privacy, they
can verify privacy for mechanisms like the Laplace and
the Exponential mechanism, and they can capture advanced examples that
go beyond mechanisms and composition, like the private vertex cover
algorithm of Gupta et al.~\cite{GLMRT10}.  
The difficulty with their approach is that it relies on a customized
and complex logic. Moreover, ad hoc rules for loops are required for
many advanced examples.

% However, the
% approach used by these system  is also 
% rather complex and difficult to use even if they have been made more practical by
% the  integration of SMT solvers to reduce the proof burden.

% \jh{I can expand here, if we want to explain these in more detail.}
% \mg{I would expand a bit by adding a short sentence comparing our approach with
%   all the ones above.}

\paragraph*{Verifying 2-safety properties}
Beyond differential privacy, there is a large body of literature on verifying
$2$-safety properties.  
Our work is most closely related to deductive methods based on program
logics;
 more precisely, approaches that reduce $2$-safety of a program $c$ to
safety of a program $c'$ built from $c$.  Such approaches include {\em
  self-composition}~\cite{BartheDR04}, {\em product programs}~\cite{ZaksP08},
and {\em type-directed product programs}~\cite{TerauchiA05}. These approaches
are subsumed by work by Barthe et al.~\cite{BartheCK11,BartheCK13}.

Another alternative is to reason directly on two programs (or two
executions of the same program) using relational program logics such
as Benton's relational Hoare logic~\cite{Benton04}, or specialized
relational logics, e.g., for information flow~\cite{AmtoftB04}. 
% {\sf  CertiCrypt}~\cite{BartheGZ09}, {\sf EasyCrypt}~\cite{BartheGHZ11},
% and {\sf EasyPriv}~\cite{BartheDGKZ13}, are computer-aided tools that
{\sf  CertiCrypt}~\cite{BartheGZ09}, and {\sf EasyCrypt}~\cite{BartheGHZ11,BartheDGKZ13}, are computer-aided tools that
support relational reasoning about probabilistic programs and have
been used to prove security of cryptographic constructions and
computational differential privacy of protocols. For such
applications, reasoning about structurally different programs is
essential.

Chaudhuri et al.~\cite{ChaudhuriGL10} develop an automated method for
analyzing the continuity and the robustness of programs. Robustness is a
$2$-safety property that is
 very similar to sensitivity as used in differential privacy.
An interesting aspect of their work is that their analysis is able to
reason about two {\em unsynchronized} pairs of executions; that is, pairs of
executions that may have different control flow.

\paragraph*{Verification of hyperproperties}
Developing general verification methods for hyperproperties
remains a challenge; however, there have been some recent proposals
in this direction (e.g.,~\cite{ClarksonFKMRS14,MilushevC13}).

\paragraph*{Other work}
There is an extensive body of work on deductive verification of
non-probabilistic and probabilistic programs, as well as many works
that consider product constructions of Labeled Transition Systems;
summarizing this large literature is beyond the scope of this paper.

\section{Conclusion}
We have proposed a program transformation that reduces proving
$(\epsilon,\delta)$-differential privacy of a probabilistic program to
proving a safety property of a deterministic transformed program. The
method applies to all standard examples where privacy is achieved
through mechanisms and composition theorems; on the other hand,
differentially private algorithms based on ad hoc output perturbation,
such as the differentially private vertex cover
algorithm~\cite{GLMRT10}, are more difficult to handle. In particular,
they fall outside the scope of Theorem~\ref{thm:hldiffpriv} which
proves the soundness of our approach. Our method is particularly
suited for reasoning about differential privacy, because the
transformed program can be analyzed with standard verification
tools. Our method can also be extended to reason about probabilistic
non-interference, at the cost of targeting an assertion language that
supports existential quantification over functions. Directions for
further work include extending the scope of
Theorem~\ref{thm:hldiffpriv} to deal with more complex examples, like
vertex cover. On a more practical side, it would be interesting to
implement our transformation for a realistic setting, for instance modeling
the \pinq language~\cite{mcsherry.pinq09}. 

\subsection*{Acknowledgments}
We thank the anonymous reviewers for their close reading and suggestions.
This research is partially supported by European project FP7-291803 AMAROUT II,
Spanish projects TIN2009-14599 DESAFIOS 10, TIN2012-39391-C04-01 Strongsoft, and
Madrid regional project S2009TIC-1465 PROMETIDOS.
Marco Gaboardi has been supported by the European Community's Seventh Framework
Programme FP7/2007-2013 under grant agreement No. 272487. Justin Hsu has been
supported by NSF grant CNS-1054229.

\bibliographystyle{plain}
\bibliography{main}

\iffull
\onecolumn
\newpage
\appendices
\section{Auxiliary lemmas}

The following is an auxiliary result used in the proof of correctness
of the method based on self-products.
\begin{lemma}
\label{prop:compseq}
Suppose that for all memories $m_1,m_2$ such that $m_1\,\Pre\,m_2$ we
have that $c$ is terminating in $m_1$ and $m_2$, and
$\glift{\Post}{(\sem{c}\,m_1)}{(\sem{c}\,m_2)}{\epsilon}{\delta}$.
Then, for every memory distributions $\mu_1,\mu_2$ such that
$\glift{\Pre}{\mu_1}{\mu_2}{\epsilon'}{\delta'}$ we have
$$\glift{\Post}
{(\sem{c}^\star\,\mu_1)}
{(\sem{c}^\star\,\mu_2)}
{\epsilon+\epsilon'}{\delta+\delta'}$$
\end{lemma}

The following is another auxiliary result used in the proof of correctness.
\medskip

\begin{lemma}
\label{lem:onedistr}
For all memories $m_1,m_2$ such that $m_1\,\Pre\,m_2$ we
have that
$\glift{\Pre}{\charfun_{m_1}}{\charfun_{m_2}}{0}{0}$.
\end{lemma}
\begin{IEEEproof}
We can take as witness $\hat{\mu}=\charfun_{m_1,m_2}$.
\end{IEEEproof}
\medskip

\begin{lemma}
\label{prop:compsequnit}
Suppose that for  $m_1,m_2$ such that $m_1\,\Pre\,m_2$ we
have that
$\glift{\Post}{(\mu_1\, m_1 )}{(\mu_2\, m_2)}{\epsilon}{\delta}$.
Then, 
$$\glift{\Post}
{((\lambda v. \charfun_{m_1\{v/x\}})^\star\,\mu_1)}
{((\lambda v. \charfun_{m_2\{v/x\}})^\star\,\mu_2)}
{\epsilon}{\delta}$$
\end{lemma}
\begin{IEEEproof}
 By Lemma \ref{lem:onedistr} and Lemma \ref{prop:compseq}. 
\end{IEEEproof}
\medskip

The proof of the next two auxiliary lemmas are presented in the work
in apRHL~\cite{BartheKOZ13}.
\begin{lemma}
  \label{lem:frame}
  Given a relation $S$ that is \emph{preserved} by $c$, i.e. such that:
  \begin{displaymath}
    \forall m_1,m_2.~ (m_1,m_2)\in S 
    \Rightarrow
    (\forall
    m_1',m_2'. (\sem{c}\,m_1\,m_1'\neq0\land\sem{c}\,m_2\,m_2'\neq0 
       \Rightarrow (m_1',m_2')\in S))
  \end{displaymath}
  If 
  \begin{displaymath}
    \forall m_1,m_2.~ (m_1,m_2)\in R \Rightarrow 
    \glift{Q}{(\sem{c}\,m_1)}{(\sem{c}\,m_2)}{\epsilon}{\delta}
  \end{displaymath}
  then
  \begin{displaymath}
    \forall m_1,m_2.~ (m_1,m_2)\in (R\cap S) \Rightarrow \glift{(Q\cap
    S)}{(\sem{c}\,m_1)}{(\sem{c}\,m_2)}{\epsilon}{\delta}
  \end{displaymath}
\end{lemma}

\begin{lemma}
  \label{lem:rand}
  For all distribution expressions $\mu_1$, $\mu_2$, if
  \begin{displaymath}
    \Delta_\epsilon(\sem{\mu_1}\,m_1,\sem{\mu_2}\,m_2)\leq \delta
  \end{displaymath}
  then 
  \begin{displaymath}
    \glift{Q}{(\sem{\Rand{x}{\mu_1}}\,m_1)}{(\sem{\Rand{x}{\mu_2}}\,m_2)}{\epsilon}{\delta}
  \end{displaymath}
  where $Q=\{(m_1,m_2)\mid m_1\,x=m_2\,x\}$. 
\end{lemma}

\section{Proof of the main theorem}

Theorem~\ref{thm:hldiffpriv} is a corollary of the following lemma:
\begin{lemma}
\label{lem:this}
\begin{displaymath}
  \HL{\Pre\land\alphav=0\land\deltav=0}{\SProd{c}}{\Post\land\alphav\leq\epsilon\land\deltav\leq\delta}
\end{displaymath}
implies
\begin{displaymath}
  \forall m_1,m_2.~ m_1\,\Pre\,m_2 \Rightarrow
  \glift{\Post}{(\sem{c}\,m_1)}{(\sem{c}\,m_2)}{\epsilon}{\delta}
\end{displaymath}

\begin{IEEEproof}
  We first introduce some new notation.  For any disjoint memories
  $m_1,m_2$ and real values $\epsilon,\delta$,
  $m_1\oplus_{\epsilon,\delta}m_2$ denotes the memory $m$ such that
  $m\,x=m_1\,x$ for every $x\in\dom(m_1)$, $m\,x=m_2\,x$ for every
  $x\in\dom(m_2)$, and $m\,\alphav=\epsilon$ and $m\,\deltav=\delta$.
  Given a memory relation $R\subseteq \Mem\times \Mem$, we let
  $\hat{R}_{\langle\epsilon,\delta\rangle}$ stand for the set
  $\{m_1\oplus_{\epsilon',\delta'}m_2\mid (m_1,m_2)\in R \land
  \epsilon'\leq\epsilon \land \delta'\leq\delta\}$.
The proof follows by structural induction on $c$, proving the following lemma:
let $R,Q\subseteq \Mem\times\Mem$ be relations on memories, then
\begin{displaymath}
\begin{array}{l}
\left(
  \forall m.~ m\in\hat{R}_{\epsilon,\delta}\Rightarrow 
      \forall m'.~ m'\in (\sem{\SProd{c}}\,m) \Rightarrow
      m'\in\hat{Q}_{\epsilon',\delta'}
\right) \\
\Longrightarrow \\
\forall m_1,m_2.~ (m_1,m_2)\in R \Rightarrow \glift{Q}{(\sem{c}\,m_1)}
   {(\sem{c}\,m_2)}{\epsilon'-\epsilon}{\delta'-\delta}
\end{array}
\end{displaymath}
Indeed, by setting $\epsilon=0$ and $\epsilon'=\epsilon$, we get the
statement of Lemma~\ref{lem:this}.  

\begin{itemize}
\item \emph{Sequential composition:} Let $(m_1,m_2)\in R$. By
  definition, $m_1\oplus_{\epsilon,\delta}m_2 \in
  \hat{R}_{\epsilon,\delta}$. Since $m'\,\alphav=m''\,\alphav$ for all
  $m',m''\in \sem{\SProd{c_1}}\,(m_1{\oplus_{\epsilon,\delta}}m_2)$,
  then there are $\epsilon_0,\delta_0$ and $S\subseteq \Mem\times\Mem$
  such that
  $\hat{S}_{\epsilon_0.\delta_0}=\sem{\SProd{c_1}}\,m_1\oplus_{\epsilon,\delta}m_2$.
  Also, from the hypotheses, for all $m\in\hat{S}_{\epsilon,\delta}$
  we have that $\forall
  m'\in(\sem{\SProd{c_2}}\,m).~m'\in\hat{Q}_{\epsilon',\delta'}$.  By
  inductive hypothesis we have thus
  \begin{enumerate}
    \item $\glift{S}{(\sem{c_1}\,m_1)}{(\sem{c_1}\,m_2)}{\epsilon_0-\epsilon}{\delta_0-\delta}$
    \item for all $m',m''$ such that $(m',m'')\in S$, we have
      $(\sem{c_2}\,m')Q_{\epsilon'-\epsilon_0,\delta'-\delta_0}(\sem{c_2}\,m'')$
  \end{enumerate}
  It follows from Lemma~\ref{prop:compseq} that
  $\glift{Q}{(\sem{c_2}^\star(\sem{c_1}\,m_1))}{(\sem{c_2}^\star(\sem{c_1}\,m_2))}{\epsilon-\epsilon}{\delta'-\delta}$.

\item \emph{While loop:}
We start by proving the following auxiliary result:
\begin{displaymath}
\begin{array}{l}
(\forall m.~m\in \hat{R}_{\epsilon,\delta} \Rightarrow
(b_1=b_2)\,m\land \hat{w}_i\,m\neq\bot\land \forall
m'\in(\hat{w}_i\,m).~m'\in \hat{Q}_{\epsilon',\delta'})
\\ \Longrightarrow \\
\forall m_1 m_2.~ (m_1,m_2)\in R\Rightarrow
\glift{Q}{(w_i\,m_1)}{(w_i\,m_2)}{\epsilon'-\epsilon}{\delta'-\delta}
\end{array}
\end{displaymath}
The proof follows by natural induction on $i$. The case $i=0$ is
trivial. For the inductive step, let $m_1,m_2\in R$. Since
$m_1\oplus_{\epsilon,\delta}m_2\in\hat{R}_{\epsilon,\delta}$, by
hypothesis we have $m_1\,b_1\Leftrightarrow m_2\,b_2$. We proceed by
case analysis on $m_1\,b_1$.
\begin{itemize}
\item In the case $\neg m_1\,b_1$, by definition of $\hat{w}_{i+1}$,
  $\hat{w}_{i+1}\,m_1\oplus_{\epsilon,\delta}m_2=m_1\oplus_{\epsilon,\delta}m_2$,
  and thus by hypothesis $m_1\oplus_{\epsilon,\delta}m_2 \in
  \hat{Q}_{\epsilon',\delta'}$, which implies $\epsilon=\epsilon'$ and
  $\delta=\delta'$. By Lemma~\ref{lem:onedistr},
  $\glift{Q}{\charfun_{m_1}}{\charfun_{m_2}}{0}{0}$, which concludes
  the proof case since we have as well $w_{i+1}\,m_1=\charfun_{m_1}$
  and $w_{i+1}\,m_2=\charfun_{m_2}$.
\item If $m_1\,b_1$ holds, then 
  \begin{displaymath}
    \hat{w}_{i+1}\,m=\bigcup_{m'\in\sem{\Seq{\SProd{c}}{\Assert{b_1\Leftrightarrow
            b_2}}}}w_i\,m' 
  \end{displaymath}
  Since $b_1\Leftrightarrow b_2$ is deterministic in
  $\sem{\SProd{c}}\,m$ and $\hat{w}_{i+1}\,m\neq\emptyset$ by
  hypothesis, then 
  \begin{displaymath}
    \hat{w}_{i+1}\,m=\bigcup_{m'\in\sem{\SProd{c}}}w_i\,m' 
  \end{displaymath}
  By the same reasoning as with sequential composition, there is then
  $S$, $\epsilon_0$, and $\delta_0$ such that
  $\hat{S}_{\epsilon,\delta}=\sem{\SProd{c}}\,m_1\oplus_{\epsilon,\delta}m_2$. Then,
  by the structural inductive hypothesis we have
  $\glift{S}{(\sem{c}\,m_1)}{(\sem{c}\,m_2)}{\epsilon_0-\epsilon}{\delta_0-\delta}$,
  and by the natural induction hypothesis 
  \begin{displaymath}
    \forall m_1 m_2.~(m_1,m_2)\in S \Rightarrow
    \glift{Q}{(w_i\,m_1)}{(w_i\,m_2)}{\epsilon'-\epsilon_0}{\delta'-\delta_0}
  \end{displaymath}
  We can conclude from Lemma~\ref{prop:compseq} that 
  \begin{displaymath}
    \glift{Q}{(w_{i+1}\,m_1)}{(w_{i+1}\,m_2)}{\epsilon'-\epsilon}{\delta'-\delta}
  \end{displaymath}
\end{itemize}

It remains to show that the property holds as well when considering
the lubs $\bigsqcup\,{\hat{w}_i}$ and $\bigsqcup\,{w_i}$:
\begin{displaymath}
\begin{array}{l}
(\forall m.~m\in \hat{R}_{\epsilon,\delta} \Rightarrow
(b_1=b_2)\,m\land \forall
m'\in(\bigsqcup\hat{w}_i\,m).~m'\in \hat{Q}_{\epsilon',\delta'})
\\ \Longrightarrow \\
\forall m_1 m_2.~ (m_1,m_2)\in R\Rightarrow
\glift{Q}{(\bigsqcup w_i\,m_1)}{(\bigsqcup w_i\,m_2)}{\epsilon'-\epsilon}{\delta'-\delta}
\end{array}
\end{displaymath}
Let $m_1$ and $m_2$ such that $(m_1,m_2)\in R$. Since
$m_1\oplus_{\epsilon,\delta}m_2$ then $\forall
m'\in(\bigsqcup\hat{w}_i\,m).~m'\in \hat{Q}_{\epsilon',\delta'}$.
Since we are considering terminating program loops, there exists $k$
such that for all $j\geq k$:
\begin{displaymath}
  \hat{w}_j (m_1\oplus{\epsilon,\delta}m_2) \neq \emptyset
\end{displaymath}
and furthermore
\begin{displaymath}
  \hat{w}_j (m_1\oplus{\epsilon,\delta}m_2) =
  \bigsqcup \hat{w}_i (m_1\oplus{\epsilon,\delta}m_2)
\end{displaymath}
From the auxiliary lemma above we have thus 
\begin{displaymath}
  \glift{Q}{(w_j m_1)}{(w_j m_2)}{\epsilon-\epsilon'}{\delta-\delta'}
\end{displaymath}
for all $j\geq k$. Since the loop termination condition is
deterministic by assumption then it also holds that $w_j m_1 =
\bigsqcup_i w_i m_1$ and $w_j m_2 = \bigsqcup_i w_i m_2$ for all
$j\geq k$. Then we can conclude:
\begin{displaymath}
  \glift{Q}{(\bigsqcup_i w_i m_1)}{(\bigsqcup_i w_i m_2)}{\epsilon-\epsilon'}{\delta-\delta'}
\end{displaymath}

%%%%%%%%%%%%%%%%%%%%%%%%%%%%%%%%%%%%%%%%%%%%
\item \emph{Laplace mechanism:}
  We consider the case $\Rand{x}{\mathsf{Lap}_\epsilon(e)}$.
  Let $m_1$ and $m_2$ such that $(m_1,m_2)\in R$. Then
  $m_1\oplus_{\epsilon_0,\delta_0}m_2\in\hat{R}_{\epsilon_0,\delta_0}$. 
  From the hypothesis
  $\sem{\SProd{c}}\,m\subseteq\hat{Q}_{\epsilon',\delta'}$ and the
  semantics of the target language, we get
  \begin{displaymath}
    \bigcup_{v\in\mathbb{R}}\;
    \left(m_1\subst{x}{v}\right)\oplus_{\epsilon_0+|\sem{e}\,m_1-\sem{e}\,m_2|\epsilon,\delta_0}\left(m_2\subst{x}{v}\right)
    \subseteq\hat{Q}_{\epsilon',\delta'}
  \end{displaymath}
  From this, we can conclude $\delta'=\delta_0$,
  $\epsilon'=\epsilon_0+|\sem{e}\,m_1-\sem{e}\,m_2|\epsilon$
  and
  \begin{displaymath}
    Q\supseteq\{(m_1,m_2)\mid \exists
    v_1,v_2.~(m_1\subst{x}{v_1},m_2\subst{x}{v_2})\in R\}
    \cap
    \{(m_1,m_2)\mid m_1\,x = m_2\,x\}
  \end{displaymath}
  Since the first term in the intersection above is preserved by any
  assignment to the $x$ variable, by Lemma~\ref{lem:frame}
  it is enough to consider the case 
  $Q=\{(m_1,m_2)\mid m_1\,x=m_2\,x\}$, and prove
  $\glift{Q}{(\sem{c}\,m_1)}{(\sem{c}\,m_2)}{|\sem{e}\,m_1-\sem{e}\,m_2|\epsilon}{0}$. 
  To verify this, by Lemma~\ref{lem:rand}, it is sufficient to show
  that
  \begin{displaymath}
    \Delta_{|\sem{e}\,m_1-\sem{e}\,m_2|\epsilon}(\mathsf{Lap}_\epsilon(\sem{e}\,m_1),\mathsf{Lap}_\epsilon(\sem{e}\,m_2)) \leq 0
  \end{displaymath}

  We need to show that for every $r$
 we have 
  $$\mathsf{Lap_\epsilon}(\sem{e}\,m_1)\,r -\mathsf{exp}(|\sem{e}\,m_1-\sem{e}\,m_2|\epsilon) \mathsf{Lap_\epsilon}(\sem{e}\,m_2)\,r\leq 0$$
  Then, it is enough to prove:
  $$ \left  (\frac{\exp{ - \frac{\epsilon|r-\sem{e}\, m_1|}{2} }}{\sum_{r'} \exp{
    -\frac{\epsilon|r'-\sem{e}\, m_1|}{2} }}\right )
 -\mathsf{exp}(|\sem{e}\,m_1-\sem{e}\,m_2|\epsilon) \left (\frac{\exp{ - \frac{\epsilon|r-\sem{e}\, m_2|}{2} }}{\sum_{r'} \exp{
    -\frac{\epsilon|r'-\sem{e}\, m_2|}{2} }}\right )\leq 0$$
This is equivalent to prove
$$
\frac{\exp{-\frac{\epsilon|r-\sem{e}\, m_1|}{2}}\cdot \sum_{r'\in\mathcal{R}}\exp{-\frac{\epsilon |r'-\sem{e}\, m_2|}{2}}}{\exp{-\frac{\epsilon
       |r-\sem{e}\, m_2|}{2}}\cdot \sum_{r'\in\mathcal{R}}\exp{-\frac{\epsilon
       |r'-\sem{e}\, m_1|}{2}}} \leq \mathsf{exp}(|\sem{e}\,m_1-\sem{e}\,m_2|\epsilon)
$$
The first term can be bound by
$$
\exp{\frac{\epsilon
       |\sem{e}\, m_2-\sem{e}\, m_1| }{2}}\cdot 
\frac{\sum_{r'\in\mathcal{R}}\exp{-\frac{\epsilon |r'-\sem{e}\, m_2|}{2}}}{\sum_{r'\in\mathcal{R}}\exp{-\frac{\epsilon
       |r'-\sem{e}\, m_1|}{2}}} 
$$
For every $r'\in \mathcal{R}$, we
know $|r'-\sem{e}\,m_2| \geq 
|r'-\sem{e}\,m_1|-|\sem{e}\,m_2-\sem{e}\,m_1|$.
So, the above can be bound by 
$$
\exp{\frac{\epsilon
       |\sem{e}\, m_2-\sem{e}\, m_1| }{2}}\cdot 
\frac{\sum_{r'\in\mathcal{R}}\exp{-\frac{\epsilon (|r'-\sem{e}\,m_1|-|\sem{e}\,m_2-\sem{e}\,m_1|)}{2}}}{\sum_{r'\in\mathcal{R}}\exp{-\frac{\epsilon
       |r'-\sem{e}\, m_1|}{2}}} 
$$
that is equivalent to
$$
\exp{\frac{\epsilon |\sem{e}\, m_2-\sem{e}\, m_1| }{2}}
\cdot 
\exp{\frac{\epsilon |\sem{e}\, m_2-\sem{e}\, m_1| }{2}}
\cdot
\frac{\sum_{r'\in\mathcal{R}}\exp{-\frac{\epsilon
       |r'-\sem{e}\, m_1|}{2}}}{\sum_{r'\in\mathcal{R}}\exp{-\frac{\epsilon
       |r'-\sem{e}\, m_1|}{2}}} 
$$
and simplifying
$$
\exp{ \frac{\epsilon |\sem{e}\, m_2-\sem{e}\, m_1| }{2}}
\cdot 
\exp{\frac{\epsilon |\sem{e}\, m_2-\sem{e}\, m_1| }{2}}
$$
that is what we need.

\item \emph{Exponential mechanism:} 
  Following a similar reasoning to
  the Laplace mechanism case, we need to prove that for every $r$ we
  have
  $$\mathsf{Exp_\epsilon}(\sem{s}\,m_1,\sem{e}\,m_1)\,r
  -\mathsf{exp}(\epsilon_1-\epsilon_0)
  \mathsf{Exp_\epsilon}(\sem{s}\,m_2,\sem{e}\,m_2)\,r\leq0$$ 
  where $\mathsf{Exp}_\epsilon(s,x)$ stands for the distribution
  $$
  \lambda r. \frac{\exp{\frac{\epsilon
        s(x,r)}{2}}}{\sum_{r'\in\mathcal{R}}\exp{\frac{\epsilon s(x,r')}{2}}}
  $$

By Lemma \ref{prop:compsequnit} and the fact that
$\sem{s}m_1=\sem{s}m_2=\hat{s}$ it is
then  enough to prove:
  $$  \frac{\exp{\frac{\epsilon
       \hat{s}(\sem{e}m_1,r)}{2}}}{\sum_{r'\in\mathcal{R}}\exp{\frac{\epsilon \hat{s}(\sem{e}m_1,r')}{2}}}-\mathsf{exp}(\epsilon_1-\epsilon_0)  \frac{\exp{\frac{\epsilon
       \hat{s}(\sem{e}m_2,r)}{2}}}{\sum_{r'\in\mathcal{R}}\exp{\frac{\epsilon \hat{s}(\sem{e}m_2,r')}{2}}})\leq 0$$
This is equivalent to prove
$$
\frac{\exp{\frac{\epsilon
       \hat{s}(\sem{e}m_1,r)}{2}}\cdot \sum_{r'\in\mathcal{R}}\exp{\frac{\epsilon \hat{s}(\sem{e}m_2,r')}{2}}}{\exp{\frac{\epsilon
       \hat{s}(\sem{e}m_2,r)}{2}}\cdot \sum_{r'\in\mathcal{R}}\exp{\frac{\epsilon
       \hat{s}(\sem{e}m_1,r')}{2}}} \leq \mathsf{exp}(\epsilon_1-\epsilon_0)
$$
Continuing we have
$$
\exp{\frac{\epsilon
       (\hat{s}(\sem{e}m_1,r)-\hat{s}(\sem{e}m_2,r)) }{2}}\cdot 
\frac{\sum_{r'\in\mathcal{R}}\exp{\frac{\epsilon \hat{s}(\sem{e}m_2,r')}{2}}}{\sum_{r'\in\mathcal{R}}\exp{\frac{\epsilon
       \hat{s}(\sem{e}m_1,r')}{2}}} \leq \mathsf{exp}(\epsilon_1-\epsilon_0)
$$
Using the fact that $\max_{r\in \mathcal{R}}|\hat{s}(e_1,r)-\hat{s}(e_2,r)|\epsilon\leq
  \epsilon_1-\epsilon_0$ we have:
$$
\exp{\frac{\epsilon_1-\epsilon_0}{2}}
\cdot 
\frac{\sum_{r'\in\mathcal{R}}\exp{\frac{\epsilon \hat{s}(\sem{e}m_2,r')}{2}}}{\sum_{r'\in\mathcal{R}}\exp{\frac{\epsilon
       \hat{s}(\sem{e}m_1,r')}{2}}} \leq \mathsf{exp}(\epsilon_1-\epsilon_0)
$$
Using the same fact we also know that for every $r'\in \mathcal{R}$ we
have $\hat{s}(e_2,r)\leq  \frac{\epsilon_1-\epsilon_0}{\epsilon}
+\hat{s}(e_1,r)$.
So,we have:
$$
\exp{\frac{\epsilon_1-\epsilon_0}{2}}
\cdot 
\frac{\sum_{r'\in\mathcal{R}}\exp{\frac{ (\epsilon_1-\epsilon_0)+
      \epsilon \hat{s}(\sem{e}m_1,r')}{2}}}{\sum_{r'\in\mathcal{R}}\exp{\frac{\epsilon
       \hat{s}(\sem{e}m_1,r')}{2}}} \leq \mathsf{exp}(\epsilon_1-\epsilon_0)
$$
that is equivalent to
$$
\exp{\frac{\epsilon_1-\epsilon_0}{2}}
\cdot 
\exp{\frac{\epsilon_1-\epsilon_0}{2}}
\cdot
\frac{\sum_{r'\in\mathcal{R}}\exp{\frac{
      \epsilon \hat{s}(\sem{e}m_1,r')}{2}}}{\sum_{r'\in\mathcal{R}}\exp{\frac{\epsilon
       \hat{s}(\sem{e}m_1,r')}{2}}} \leq \mathsf{exp}(\epsilon_1-\epsilon_0)
$$
and simplifying
$$
\exp{\frac{\epsilon_1-\epsilon_0}{2}}
\cdot 
\exp{\frac{\epsilon_1-\epsilon_0}{2}}
\leq \mathsf{exp}(\epsilon_1-\epsilon_0)
$$

\end{itemize}
\end{IEEEproof}
\end{lemma}

\begin{lemma}[Proof of the accuracy specification]
\label{lem:secondSpecPTR}

\begin{displaymath}
\begin{array}{l}
\forall m_1,m_2.~ \sem{e}\,m_1=\sem{e}\,m_2 \Rightarrow \glift{Q}{(\sem{\Ass{x}{\mathsf{Lap}_\epsilon(e)}}\,m_1)}
   {(\sem{\Ass{x}{\mathsf{Lap}_\epsilon(e)}}\,m_2)}{0}{\delta}
\end{array}
\end{displaymath}
where
\begin{displaymath}
  Q \doteq \{(m_1,m_2)\mid m_1\,x =m_2\,x \land |m_1\,x-\sem{e}\,m_1|\leq \log(2/\delta)/(2\epsilon)\} 
\end{displaymath}
\end{lemma}

\begin{IEEEproof}
  We need to prove:
$$ \glift{Q}{(\sem{\Rand{x}{\mathsf{Lap_\epsilon}(e)}}\,m_1)}{(\sem{\Rand{x}{\mathsf{Lap_\epsilon}(e)}}\,m_2)}{0}{\delta}$$
By the assumption $e_1=e_2$ we have that
$\sem{\Rand{x}{\mathsf{Lap_\epsilon}(e)}}\,m_1$ and
$\sem{\Rand{x}{\mathsf{Lap_\epsilon}(e)}}\,m_2$ are the same
distribution $\hat{\mu}$.
Now, consider the set $S = \{ z : \mathbb{R} \mid |z -
  \sem{e_1}\,m_1| < \log(2/\delta)/(2\epsilon)\}$ and the 
  distribution $\mu\in\distr(\mathbb{R}\times \mathbb{R})$, parametrized on
  $S$ defined as:
  \begin{equation*}
    \mu(z_1, z_2) := \left\{
      \begin{array}{ll}
\hat{\mu}\, z_1 &\text{if
        } z_1 = z_2 \land z_1\in S\\
        0 &\text{otherwise.}
      \end{array}
    \right.
  \end{equation*}
Notice that by definition of $\hat{\mu}$ we have
$\pi_1\mu\leq \sem{\Rand{x}{\mathsf{Lap_\epsilon}(e)}}{m_1}$ and 
$\pi_2\mu\leq \sem{\Rand{x}{\mathsf{Lap_\epsilon}(e)}}{m_2}$.
Moreover, by definition of $S$ we also have that for every $m$, $\mu\, m\neq 0\Rightarrow 
\Post\, m$.
Since clearly $\pi_1\mu=\pi_2\mu$, the only thing left to prove is
that   $\Delta_{0}(\hat{\mu}, \pi_1\mu)\leq \delta$. This means that
we need to prove
$$
\max_{R\subseteq\mathbb{R}}\{ \hat{\mu}\,R-\pi_1\mu\,R\}\leq \delta
$$
It is easy to see that on values in $\mathbb{R}\cap S$ the two distribution
coincide. So we can instead consider
$$
\max_{R\subseteq(\mathbb{R}/ S)}\{ \hat{\mu}\,R-\pi_1\mu\,R\}\leq \delta
$$
Now, notice that for every $R\subseteq(\mathbb{R}/ S)$ we have
$\pi_1\mu=0$, so we can just consider
$$
\max_{R\subseteq(\mathbb{R}/ S)}\{ \hat{\mu}\,R\}\leq \delta
$$
and since by definition $\hat{\mu}\,R=\sum_{a\in R}\hat{\mu}\,a$ where
every value is non-negative, we can just consider  $\hat{\mu}\,(\mathbb{R}/S)\leq \delta$.
Now, recall that $S$ corresponds to the interval:
$$
[-(\log(2/\delta)/(2\epsilon)) + \sem{e}\,m_1, \sem{e}\,m_1 +(\log(1/\delta)/\epsilon)]
$$ 
and that $\hat{\mu}=\sem{\Rand{x}{\mathsf{Lap_\epsilon}(e)}}\,m_1$. So, we can apply a tail
  bound on the Laplace distribution:
  \[
    \{\hat{\mu}\,z \ |\ z \in (\mathbb{R}/S)\} = \{\hat{\mu}\,z \ |\ |z - \sem{e}\,m_1| \geq \log(2/\delta)/(2\epsilon)\} < \delta.
  \]
and conclude
$$
\hat{\mu}\,(\mathbb{R}/S)\leq \delta
$$
that is what we need.
  \end{IEEEproof}

\medskip

\begin{lemma}[Tail bound for the discrete version of Laplace]
  Let $x$ be drawn from the discrete version of the Laplace distribution with mean $0$ and parameter
  $b > 0$, i.e., with probability 
  \[
    L_{b}(x) = \frac{\exp{ -\frac{|x|}{2b} }}{\sum_{z\in \mathbb{Z}} \exp{ -\frac{|z|}{2b} }}.
  \]
  Then, for $T \in \mathbb{N}$ and $T > 0$:
  \[
    \Pr{x}{|L_{b}(x)| > T }\leq 2\,\exp{ -\frac{T}{2b} }.
  \]
  In particular, if $b = 1/\epsilon$ (like in $\nm{Lap}_\epsilon(x)$) and $T =
  \log(2/\delta)/(2\epsilon)$, we have Lemma~\ref{lem:accuracy}.
\end{lemma}
\begin{IEEEproof}
  We have
  \begin{align*}
    \Pr{x}{|L_{b}(x)| > T }  &=    \Pr{x}{\left|\frac{\exp{ -\frac{|x|}{2b} }}{\sum_{z\in \mathbb{Z}} \exp{
          -\frac{|z|}{2b} }}\right| > T }\\
&=    \Pr{x}{\frac{\exp{ -\frac{|x|}{2b} }}{\sum_{z\in \mathbb{Z}}
    \exp{ -\frac{|z|}{2b} }} > T }+\Pr{x}{\frac{\exp{ -\frac{|x|}{2b} }}{\sum_{z\in \mathbb{Z}} \exp{ -\frac{|z|}{2b}
    }} < -T }\\
&=    \sum_{x=T}^{\infty}\frac{\exp{ -\frac{|x|}{2b} }}{\sum_{z\in \mathbb{Z}}
    \exp{ -\frac{|z|}{2b} }} +\sum_{-T}^{-\infty}\frac{\exp{ -\frac{|x|}{2b} }}{\sum_{z\in \mathbb{Z}} \exp{ -\frac{|z|}{2b}
    }}\\
&=    \frac{\sum_{x=T}^{\infty}\exp{ -\frac{x}{2b} }+\sum_{-T}^{-\infty}\exp{ \frac{x}{2b} }}{\sum_{z\in \mathbb{Z}} \exp{ -\frac{|z|}{2b}}}\\
&=    \frac{\sum_{x=0}^{\infty}\exp{ -\frac{x+T}{2b}
  }+\sum_{x=0}^{\infty}\exp{ \frac{-x-T}{2b} }}{\sum_{z\in \mathbb{Z}}
  \exp{ -\frac{|z|}{2b}
    }}\\
&=    \frac{\exp{-\frac{T}{2b}}\,\sum_{x=0}^{\infty}\exp{ -\frac{x}{2b}
  }+\exp{-\frac{T}{2b}}\, \left (1+\sum_{x=1}^{\infty}\exp{
      -\frac{x}{2b} }\right )}{\sum_{z\in \mathbb{Z}}
  \exp{ -\frac{|z|}{2b}
    }}\\
&=    \exp{-\frac{T}{2b}}\,\frac{\left ( \sum_{x=0}^{\infty}\exp{ -\frac{x}{2b}
  } +1+\sum_{x=1}^{\infty}\exp{
      -\frac{x}{2b} }\right )}{\sum_{z\in \mathbb{Z}}
  \exp{ -\frac{|z|}{2b}
    }}\\
&=    \exp{-\frac{T}{2b}}\,\left (\frac{1}{\sum_{z\in \mathbb{Z}}
  \exp{ -\frac{|z|}{2b}}} +\frac{\left ( \sum_{x=0}^{\infty}\exp{ -\frac{x}{2b}
  } +\sum_{x=1}^{\infty}\exp{
      -\frac{x}{2b} }\right )}{\sum_{z\in \mathbb{Z}}
  \exp{ -\frac{|z|}{2b}}} \right)\\
&=    \exp{-\frac{T}{2b}}\,\left (\frac{1}{\sum_{z\in \mathbb{Z}}
  \exp{ -\frac{|z|}{2b}}} +1\right )\\
&\leq  2\, \exp{-\frac{T}{2b}}
\end{align*}
\end{IEEEproof}

\section{Verification of vertex cover}\label{app:vertex-cover}
The extended logic used to prove the vertex cover in apRHL features a
more precise rule for while loops, that allows the privacy budget to
vary at each iteration
\begin{displaymath}
\infrule
{\begin{array}{@{}l@{}}
 \Inv \implies 
 b_1 \equiv b_2 \land i_1 = i_2
 \qquad
   \Inv \land n \leq i_1 \implies \lnot b_1 \\
   \AEquiv{\Inv \land b_1 \land i_1 = j}
   {c_1}
   {c_2}
   {\Inv \land i_1 = j \!+\! 1}
   {\epsilon_j}{\delta_j}
\end{array}}
{\AEquiv{\Inv \land i_1 = 0}
        {\While{b_1}{c_1}}
        {\While{b_2}{c_2}}
        {\Inv \land \lnot b_1}
        {\sum_{i=0}^{n-1} \epsilon_i}{\sum_{i=0}^{n-1}  \delta_i}}
\end{displaymath}
and a code motion rule that allows to swap statements $c_1$ and $c_2$
provided they satisfy some independence condition:
$$
  \AEquiv{\forall x\in X. x_1=x_2}{c_1;c_2}{c_2;c_1}{\forall x\in X.x_1=x_2}{0}{0}
$$ 
In addition, the extended logic features a transitivity rule that
allows to compose apRHL judgments.  These rules can be readily encoded
in our setting, provided we allow for more general forms of products
as considered in \cite{BartheCK11,BartheCK13}.

However, the extended logic also considers a probabilistic programming
language with $\mathsf{assert}$ statements, and ad hoc rules for
random assignments and while loops:
\begin{displaymath}
\infrule
{\begin{array}{@{}l@{}}
 \Inv \implies 
 b_1 \equiv b_2 \land P_1 \equiv P_2  \\
\AEquiv{\Inv \land b_1 \land \lnot P_1}
        {c_1;\, \Assert{P_1}}
        {c_2;\, \Assert{P_2}}
        {\Inv}
        {\epsilon}{\delta} \\
\AEquiv{\Inv \land b_1}
   {c_1}
   {c_2}
   {\Inv}
   {0}{0} \\
\AEquiv{\Inv \land b_1 \land P_1}
        {c_1}
        {c_2}
        {\Inv \land P_1}
        {0}{0}
\end{array}}
{\AEquiv{\Inv}
        {\While{b_1}{c_1}}
        {\While{b_2}{c_2}}
        {\Inv \land \lnot b_1}
        {\epsilon}{\delta}}
\end{displaymath}
These rules are not captured by our approach.

For comparison, we briefly describe the proof in apRHL and the
relational specifications of \textsf{choose} that are required for
completing the proof.  For the first case, the {\sf apRHL} proof uses
the first generalized loop rule, and the following property of
$\mathsf{choose}$:
$$\AEquiv{g_1 \cup\{\langle{u_1,t_1}\rangle\} =g_2}{
\Ass{v_1}{\mathsf{choose}_{\epsilon,n}(g_1)}}{\Ass{v_2}{\mathsf{choose}_{\epsilon,n}(g_2)}}{v_1=v_2}{\epsilon/\left(2\sqrt{n}\sqrt{|g_1|}\right)}{0}
$$ 
In the second case, the {\sf apRHL} uses the second generalized loop rule,
and the following properties of $\mathsf{choose}$:

\begin{align*}
\AEquiv{g_1 =g_2\cup\{\langle{u_2,t_2}\rangle\}}{
  \left
      (\begin{array}{c}\Ass{v_1}{\mathsf{choose}_{\epsilon,n}(g_1)};\\\Assert{v_1\neq
        u_1\wedge v_1\neq t_1}\end{array}\right )}{
\left
      (\begin{array}{c}\Ass{v_2}{\mathsf{choose}_{\epsilon,n}(g_2)};\\
        \Assert{v_2\neq u_2\wedge v_2\neq t_2}\end{array}\right )}{
 v_1 =v_2}{0}{0}, 
\\
\\
\AEquiv{g_1 =g_2\cup\{\langle{u_2,t_2}\rangle\}}{
\left
      (\begin{array}{c}\Ass{v_1}{\mathsf{choose}_{\epsilon,n}(g_1)};\\
        \Assert{v_1= u_1\vee v_1=t_1}\end{array}\right )}
{\left
      (\begin{array}{c}\Ass{v_2}{\mathsf{choose}_{\epsilon,n}(g_2)};\\
        \Assert{v_2=q u_2\vee v_2 t_2}\end{array}\right )}{v_1 =v_2}{\frac{\epsilon}{4}}{0}.
\end{align*}
\fi
\end{document}